\documentclass[longauth]{aa}  

\usepackage{graphicx}
\usepackage{txfonts}
\usepackage{longtable}
\usepackage{supertabular,booktabs}
\usepackage{tabularray}
\usepackage{subfig}
\usepackage{pdflscape}
\usepackage{lipsum}
\usepackage{xcolor}
\usepackage{threeparttable}

\newcommand{\CellWithForecedBreak}[2][c]{
\begin{tabular}[#1]{@{}c@{}}#2\end{tabular}}
\usepackage{multirow}
\usepackage[]{hyperref}
\hypersetup{
    colorlinks=true,
    citecolor=blue,
    linkcolor=blue,
    urlcolor=blue,
    }
\begin{document}

   \title{Light-Curve and Spectral Properties of Type II Supernovae from the ATLAS survey}

   \author{K.~Ertini
          \inst{1,2}
          \and
          J.~P. Anderson\inst{3} 
          \and
          G. Folatelli\inst{1,2,4} 
          \and
          S. González-Gaitán\inst{5,3}
          \and
          C. P. Guti\'errez\inst{6,7}
          \and
          J. Sollerman\inst{8}
          \and
          O. Rodr\'iguez\inst{9,10}
          \and
          A. Aryan\inst{11}
          \and
          T.-W. Chen\inst{11}
          \and
          E. Concepcion\inst{12}
          \and
          S.P. Cosentino\inst{13}
          \and
          M. Dennefeld\inst{14}
          \and
          N. Erasmus\inst{15,16}
          \and
          M. Fraser\inst{17}
          \and
          L. Galbany\inst{6,7}
          \and
          M. Gromadzki\inst{18}
          \and
          C. Inserra\inst{19}
          \and
          T. E. Müller-Bravo\inst{20,21}
          \and
          P. J. Pessi\inst{8,22}
          \and
          T. Pessi\inst{3}
          \and
          T. Petrushevska\inst{12}
          \and
          G. Pignata\inst{23}
          \and
          F. Ragosta\inst{24,25}
          \and
          S. Srivastav\inst{26}
          \and
          D. R. Young\inst{27}
          }

   \institute{Facultad de Ciencias Astronómicas y Geofísicas, Universidad Nacional de La Plata, Paseo del Bosque S/N, B1900FWA, La Plata, Argentina\\
              \email{keilaertini@gmail.com}
            \and
            Instituto de Astrofísica de La Plata (IALP), CCT-CONICET-UNLP, Paseo del Bosque S/N, B1900FWA, La Plata, Argentina 
            \and
            European Southern Observatory, Alonso de C\'ordova 3107, Casilla 19, Santiago, Chile
            \and
            Kavli Institute for the Physics and Mathematics of the Universe (WPI), The University of Tokyo Institutes for Advanced Study, The University of Tokyo, Kashiwa, 277-8583 Chiba, Japan
            \and
            Instituto de Astrofísica e Ciências do Espaço, Faculdade de Ciências, Universidade de Lisboa, Ed. C8, Campo Grande, 1749-016 Lisbon, Portugal
            \and
            Institut d'Estudis Espacials de Catalunya (IEEC), Edifici RDIT, Campus UPC, 08860 Castelldefels (Barcelona), Spain
            \and
            Institute of Space Sciences (ICE, CSIC), Campus UAB, Carrer de Can Magrans, s/n, E-08193 Barcelona, Spain
            \and
           The Oskar Klein Centre, Department of Astronomy, Stockholm University, Albanova University Center, SE 106 91 Stockholm, Sweden
            \and
           Pontificia Universidad Cat\'olica de Chile, Vicu\~na Mackenna 4860, Macul, Santiago, Chile
            \and
            Instituto Milenio de Astrof\'isica (MAS), Nuncio Monse\~nor S\'otero Sanz 100, Of. 104, Santiago, Chile
            \and
            Graduate Institute of Astronomy, National Central University, 300 Jhongda Road, 32001 Jhongli, Taiwan
            \and       
           Center for Astrophysics and Cosmology, University of Nova Gorica, Vipavska 11c, 5270 Ajdov\v{s}\v{c}ina, Slovenia
            \and
           University of Catania, Department of Physics and Astronomy ``E. Majorana'', Italy
            \and
            Institut d’Astrophysique de Paris (IAP), Sorbonne Université, CNRS, Paris, France
            \and
            South African Astronomical Observatory, Cape Town, 7925, South Africa
            \and
           Department of Physics, Stellenbosch University, Stellenbosch, 7602, South Africa
            \and
            School of Physics, University College Dublin, LMI Main Building, Beech Hill Road, Dublin 4, Ireland
            \and
            Astronomical Observatory, University of Warsaw, Al. Ujazdowskie 4, 00-478 Warszawa, Poland
            \and
            Cardiff Hub for Astrophysics Research and Technology, School of Physics \& Astronomy, Cardiff University, Queens Buildings, The Parade, Cardiff, CF24 3AA, UK
            \and
           School of Physics, Trinity College Dublin, The University of Dublin, Dublin 2, Ireland
            \and
            Instituto de Ciencias Exactas y Naturales (ICEN), Universidad Arturo Prat, Chile
            \and
            Astrophysics Division, National Centre for Nuclear Research, Pasteura 7, 02-093 Warsaw, Poland
            \and
            Instituto de Alta Investigaci\'on, Universidad de Tarapac\'a, Casilla 7D, Arica, Chile
            \and
            Dipartimento di Fisica ``Ettore Pancini'', Università di Napoli Federico II, Via Cinthia 9, 80126 Naples, Italy 
            \and
            INAF - Osservatorio Astronomico di Capodimonte, Via Moiariello 16, I-80131 Naples, Italy
            \and
            Astrophysics sub-Department, Department of Physics, University of Oxford, Keble Road, Oxford, OX1 3RH, UK
            \and
            Astrophysics Research Centre, School of Mathematics and Physics, Queen’s University Belfast, Belfast BT7 1NN, UK
             }
   \date{Received XXX; accepted XXX}

 
  \abstract
   {Type II supernovae (SNe) are the most common terminal stellar explosions in the Universe. With SNe now being detected within days after explosion, there is growing evidence that the majority of Type II SNe (SNe~II) show signs of interaction with a confined, dense cirumstellar material (CSM) in the first few days post explosion.}
   {In this work we aim to bridge the gap between single SN studies showing early-time interaction in their spectra, and the statistical studies of early-time SN light curves, which imply the existence of CSM.}
   {We present a sample of 68 Type II SNe with both early photometric data, obtained with the ATLAS survey, and spectroscopic data, obtained with the ePESSTO+ collaboration. A subset of the sample is classified based on the presence or absence of narrow spectral features with electron-scattered broadened wings in the early spectra, indicative of interaction with CSM. We characterise the photometric and spectroscopic properties of the sample by measuring rise times to maximum light, peak magnitudes, decline rates and line velocities. Additionally, we measure the ratio of absorption to emission (a/e) of the $\mathrm{H\alpha}$ P-Cygni profile.}
   {Our analysis reveals that SNe II showing early spectroscopic signs of interaction with CSM decline faster and are brighter than those without. However no difference is found in rise times between the two groups. A clear separation is observed in the a/e ratio: SNe with signs of interaction exhibit lower a/e ratios at all epochs compared to those without. Our results highlight that understanding SN~II ejecta-CSM interaction requires large, uniform samples of photometric and spectroscopic data, such as the one presented in this work.
   }
   {}


   \maketitle
%

\section{Introduction}
Type II supernovae (SNe~II) are the most abundant terminal explosions, arising from stars more massive than $\approx$ 8 $M_{\odot}$ \citep{li11,shivvers17}, which have retained their hydrogen envelopes. Direct detection of progenitor stars has established conclusive evidence that these SNe arise from red supergiant (RSG) stars  \citep[e.g.][]{maund05,fraser12,vandyk12,smartt15a,kilpatrick23}. A subclass of SNe~II are Type IIn, which show narrow emission lines in their spectra. Such narrow emission lines are thought to result from the collision between the SN ejecta and slow-moving dense circumstellar material (CSM) near the progenitor star \citep{schlegel90}. Until recently, it was believed that `normal' SNe~II explode in relatively clean environments, where any interaction between the SN ejecta and their environment had negligible effect on the observed transient. However, this picture has changed during the past decade.

In recent years, there has been an increasing number of early-time observations of SNe~II, within days or even hours after the explosion. Several surveys now focus on discovering transients with high cadence, providing such early-time observations. 
Observations of some SNe~II within hours to a few days post-explosion reveal spectra dominated by narrow, highly ionized emission lines with electron-scattered wings (also called "flash features"), implying strong interaction with a dense, slow-moving CSM \citep[e.g.][]{niemela85,leonard00,yaron17}. However, in contrast to SNe~IIn, these features then gradually disappear on timescales of hours to days, and spectra transform to those of ``normal'' SNe~II. Sample studies centered on these high ionization features can be found in \citet{khazov16}, \citet{bruch21}, \citet{bruch23} , and \citet{jacobsongalan24}. 

In the last years, the discovery of two nearby SNe~II made it possible to constrain CSM properties with significant precision. SN~2023ixf and SN~2024ggi were discovered by \citet{itagaki23} and \citet{srivastav24}, respectively, and they were both located at $\sim$7~Mpc. Their early detection enabled a detailed tracking of the flash features, allowing estimations of their mass-loss ($\dot M$) histories and progenitor properties, and providing constraints on late-time stellar evolution and progenitor density profiles. Another relevant example is SN~2021yja, which was detected only $\approx$ 5 hours after explosion \citep{smith21}. Unlike SN~2023ixf and SN~2024ggi, it did not exhibit flash features in its early spectra. However, it did show broader features and a particular ``ledge-shaped'' feature around 450–480 nm, which has been interpreted as a possible sign of interaction with a low mass CSM \citep{hosseinzadeh22,soumagnac20}. These cases highlight the diversity in early-time spectroscopic signatures among SNe~II and also how the varying degrees of CSM interaction can shape early-time transient evolution.

SN-II spectra during the plateau phase also exhibit signs of interaction with CSM. For instance, \citet{faran14} analysed a sample of 23 SNe and identified potential evidence of CSM interaction in the spectra of at least six events. These SNe displayed high-velocity features in the blue wing of $\mathrm{H\alpha}$ during the plateau phase, with a similar behaviour observed in $\mathrm{H\beta}$. Subsequently, \citet{gutierrez17} conducted a comprehensive study of over 100 SNe~II and reported further potential indications of CSM interaction in a significant fraction of the sample during the recombination phase.

The early rise of SN~II light curves (LCs) can provide constraints on the progenitor and its environment. In a clean environment, the properties of the LC during the early radiation-dominated phase primarily depend on the progenitor radius, explosion energy, and progenitor mass. When the CSM is considered, the influence of the progenitor properties on the early LC diminishes, whereas the properties of the CSM have a significant effect \citep{moriya11,dessart17,dessart23,cosentino25}. Sample studies using observations of SNe~II focusing on the early behaviour elucidate the observation of this effect. The rise times in observed broad-band LCs are found to be short relative to what is expected in clean RSG progenitor explosions. This discrepancy  was first identified observationally by \citet{gall15} and \citet{gonzalezgaitan15} and later by \citet{rubin16} and \citet{bruch23}. However, these rise times are long compared with the expected shock breakout (SBO) from the stellar surface. \citet{gonzalezgaitan15} attributed the short rise times either to the shock cooling caused by the core collapse of a RSG with compact and dense envelopes, or to the delayed SBO following the collapse of a RSG with an extended atmosphere or pre-SN CSM. By analyzing a sample of 26 SNe~II, \citet{forster18} argued that the observed emission is consistent with the SBO through a dense CSM rather than directly from the stellar surface. This finding suggests that the characteristic hour-long SBO timescales expected from stellar envelopes may be rare in nature. Instead, in most SNe~II, the SBO could be delayed due to the presence of significant CSM.

In the last decade, modelling efforts have been made to understand the early behaviour of both LCs and spectra of SNe~II, and to understand how they are affected by CSM interaction. Studies that modelled RSG star explosions embedded in a variety of dense envelopes showed that the early luminosity, color evolution, and photospheric velocities are strongly affected by the properties of the CSM and that if the CSM is dense enough, the SBO is delayed \citep{moriya11,morozova17,dessart17}. When comparing some of these models with observations of SNe~II with early interaction (i.e. those showing flash ionization lines), usually high mass loss rates are required $\dot{M} \sim 10^{-3}-10^{-2} M_{\odot}~yr^{-1}$, while what is observed in RSG is on the order of $\dot{M} \sim 10^{-6} M_{\odot}~yr^{-1}$ \citep{beasor18}. However, when taking the RSG wind acceleration into account, even if the mass-loss rate of the progenitor is relatively low it can produce a dense CSM at the immediate stellar vicinity and the early LCs of SNe II can be significantly affected \citep{moriya18,moriya23}.

Discrepancies arise when comparing what is expected from standard stellar evolution theory with the observations of SNe~II with early interaction. While models predict that the luminosity and color evolution should be strongly influenced by the properties of the CSM at early times, \citet{bruch23} argued that SNe~II showing flash features are not significantly brighter, nor bluer, nor more slowly rising than those without. They suggest that the CSM creating the flash features is not massive enough to contribute significantly to the luminosity of SNe~II. On the contrary, \citet{jacobsongalan24} and \citet{jacobsongalan25} recently reported that SNe~II with early interaction do appear to be bluer and more luminous. These inconsistencies could be due to how the samples are constructed and classified based on flash spectroscopy, but they highlight the ongoing uncertainty regarding how RSG progenitors generate the required dense CSM and how this affects the observed properties of SNe~II.

The objective of this work is to combine the detailed spectroscopic analysis of individual events with large-scale statistical LC analysis to further understand the frequency and effects of early interaction on the properties of SNe~II. In this study, we present an early-time photometric analysis of a large SN~II sample. This is complemented by a large set of spectroscopic data obtained from early times to plateau epochs.

This paper is organised as follows: in Section 2 we present the sample, including the LCs and spectra. In Section 3 we outline the methods employed for the photometric and spectral measurements. Section 4 presents the correlations between parameters and in Section 5 we discuss our results.

\section{Data sample}
\label{sec:sample}
The SN~II sample was compiled between 2019 and 2023 making use of the Asteroid Terrestrial-impact Last Alert System (ATLAS; \citealp[]{tonry18,smith20}) for the photometry, and of the Public European Southern Observatory Spectroscopic Survey of Transient Objects (ePESSTO+) collaboration \citep[]{smartt15} for the spectral data. 
SNe were selected based on the following criteria: 1) being classified as a SN~II, 2) having ATLAS photometry, 3) having a constraining non-detection less than four days before the first detection, 4) being brighter than 18th mag at optical wavelengths at maximum light, and 5) being observable from La Silla Observatory and continue to be for the next few months. Taking these criteria into account, we collected data from 77 SNe. We excluded Type IIn and IIb SNe.

\begin{figure}
	\includegraphics[width=\columnwidth]{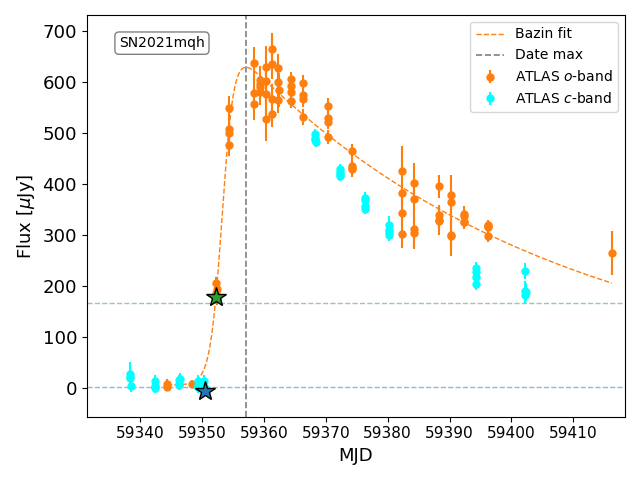}
    \caption{Example of LC of a SN II from our sample. The weighted average detection epoch and last non-detection are marked as green and blue stars, respectively. Dashed horizontal lines indicate the flux levels of the last non-detection plus one sigma uncertainty (blue line) and of the first detection minus one sigma uncertainty (green line). If the difference between the two dashed lines is > 0, we consider the non detection to be deep enough. The orange dashed line corresponds to the Bazin fit performed on the $o$-band LC (see Section~\ref{sec:bazin}), and the dashed vertical gray line shows the time of maximum light based on the fit.}
    \label{fig:example_lc}
\end{figure}

We downloaded the photometry from the ATLAS forced photometry server\footnote{\url{https://fallingstar-data.com/forcedphot/}}. The measurements from the photometry server come from a PSF fitting routine, based on algorithms from \citet{tonry11} and \citet{sonnett13}. The output gives the MJD of the start of the exposure, the AB magnitude, the flux, and the reduced $\chi^{2}$ of the PSF fit, among other parameters. Photometry is obtained in the $c$ (wavelength range 4200–6500~{\AA}) and $o$ ($5600$–$8200$~{\AA}) ATLAS passbands. 
An example of a light curve of a SN in our sample is shown in Figure \ref{fig:example_lc}.

Once a SN was selected to be a part of the sample, spectra were obtained with the ESO Faint Object Spectrograph and Camera (EFOSC2; \citealp[]{buzzoni84}) mounted on the NTT at La Silla Observatory. Our primary choice was to perform the observations with grism \#13, which covers a wavelength range of 3685-9315~{\AA} and has a resolution of 21.2~{\AA}. Occasionally, we used grisms \#11 and \#16, which cover ranges of 3380-7520~{\AA} and 6015-10320~{\AA}, with resolutions of 15.8 and 16.0~{\AA}, respectively. We typically used a 1".0 slit, but in bad seeing conditions a 1".5 slit was adopted. The integration times varied between 1500 and 3600~s. All spectra were reduced following standard procedures using the ePESSTO+\footnote{\url{https://github.com/svalenti/pessto}} pipeline. We collected a total of 194 spectra. In some SNe the Type II classification was unclear (even if such a classification had been initially reported) due to non-standard features in their spectra. In these cases we re-classified the events using the Supernova Identification code \citep[SNID;][]{blondin07}. 
We reclassified eight SNe. Five were reclassified as Type IIb: SN~2019sox, SN~2019tua, SN~2020utf, SN~2021fox, and SN~2022aczs; two Type Ia: SN~2020skx and SN~2022yeg; and one Type IIn, SN~2020ntl. We removed these SNe from our sample, but their photometry and spectral sequences will be made available via Zenodo\footnote{\url{https://zenodo.org/}}. 
Also, we note that the spectra of SN~2020uwl do not show any typical P-Cygni feature, which led us to believe that this is a peculiar event, hence we removed it from the sample, but we show its photometric and spectral behaviour in Zenodo. We identified four SNe which have an SN~1987A-like light curve behaviour: SN~2021aatd, SN~2021adcw, SN~2021adyl and SN~2022acrv. While such SNe are clearly different from `normal' SNe~II, we kept them in our sample to highlight their properties, but removed them/treat them differently in some specific analysis/correlations presented below. 

\begin{figure}
	\includegraphics[width=\columnwidth]{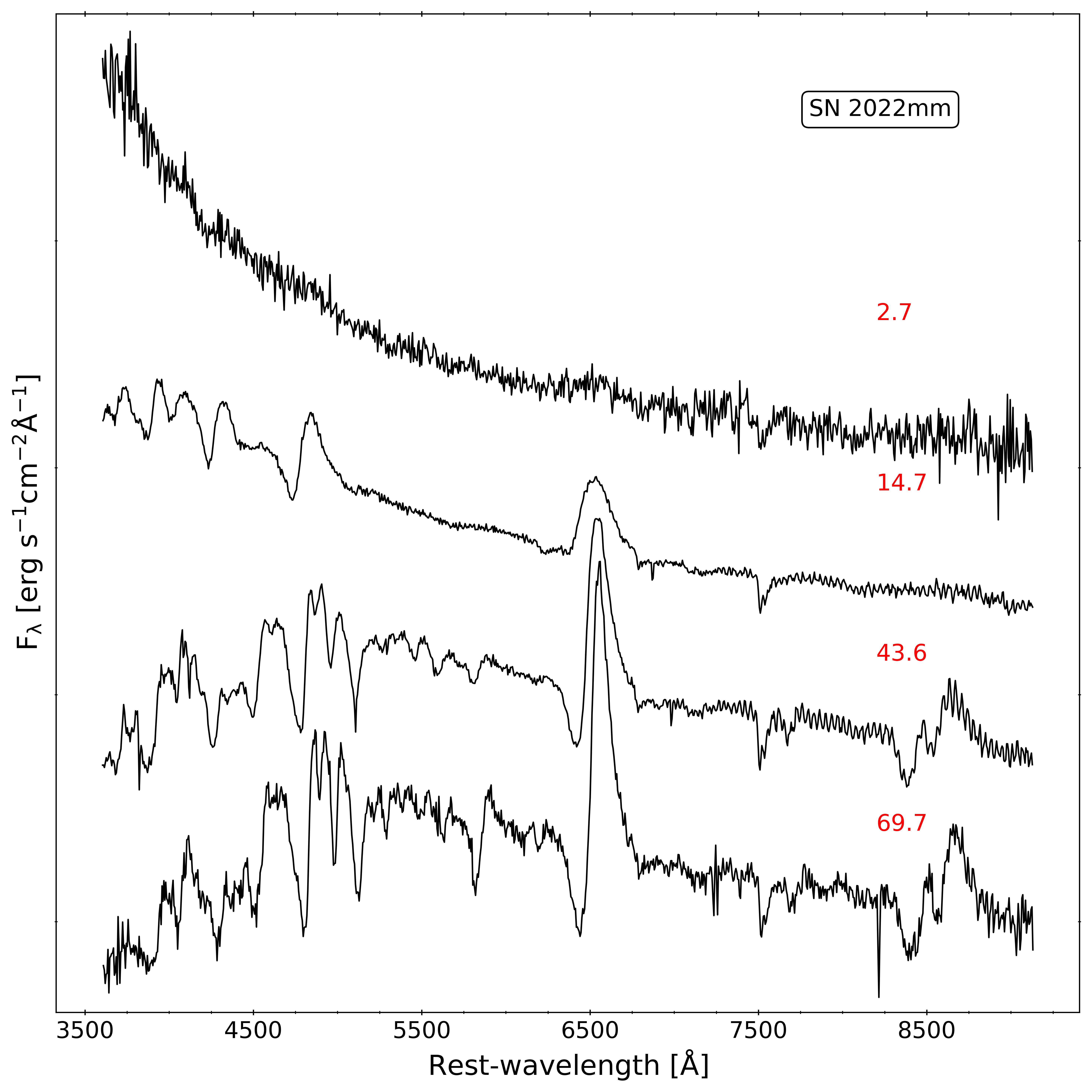}
    \caption{Example of spectral sequence for SN~2022mm. The phase in days from first light is indicated in red next to each spectrum.}
    \label{fig:spec_example}
\end{figure}

Among the remaining 68 SNe~II, there were 14 where we could not obtain any spectra. We kept these SNe as part of the sample, and we included them in all the photometric calculations, but they were not included in the spectral measurements. While we were unable to obtain any spectra with EFOSC2, each of these SNe have a classification spectrum, so we are certain they are Type IIs. Additionally, to ensure there were no bias in the selection of objects, we checked if the magnitudes of the 14 SNe for which we did not get spectra were particularly sub-luminous. We found that 13/14 SNe fall within the mean $M_{max}$ (see Section~\ref{sec:lc_dist}) and only one SN is sub-luminous, so we consider this not to be a bias. An example of a spectral sequence of an SN of our sample is shown in Figure \ref{fig:spec_example}.

The 68 selected SNe are listed in Table \ref{tab:sample}, along with their host galaxy information, and discovery date. Additionally, in Table~\ref{tab:criteria} the selection steps are listed.

\section{Light-curve and spectral measurements} 
Using $o$- and $c$-band photometry for 68 SNe~II, along with optical-wavelength spectroscopy for a subset of 54 SNe within that initial sample, this study aims to investigate the LC and spectral properties of a statistically significant sample of SNe~II with early-time photometric data. These data are then used to constrain the progenitor and explosion properties of SNe~II, and more importantly to assess the presence and effect of any significant CSM material surrounding the progenitor stars at the time of explosion. 

\begin{table}
\centering
	\caption{Initial sample size and its reduction in each step of our analysis.}
	\label{tab:criteria}
	\begin{tabular}{lccc} 
             \hline
		\multicolumn{4}{|c|}{Photometric sample}  \\
		\hline
		Step & Criteria & Removed & Objects left   \\
		\hline
            1 & Initial sample & $-$ & 77 \\
            2 & Re-classification & 9 & 68 \\
            3 & No deep non-detection & 4 & 64 \\
    	\hline
        \multicolumn{4}{|c|}{Spectroscopic sample}  \\
		\hline
		Step & Criteria & Removed & Objects left   \\
             1 & Initial sample & $-$ & 54 \\
             2 & Only photospheric spectra & 9 & 45 \\
             3 & Measurable $\mathrm{H\beta}$ & 24 & 21 \\
		\hline
	\end{tabular}
\end{table}

Following the above goals, in the below subsections we define our LC and spectral measurements to be used in characterising this SN~II sample. First, we present our first-light epoch estimation method and then outline the adopted distances and Milky-Way line-of-sight extinction (host-galaxy extinction is ignored, see our justification in Section \ref{sec:ext}). Then we define our light-curve measurements: absolute peak magnitudes, rise times to maximum light, and post-peak decline rates. Lastly, we define our spectral measurements: expansion velocities and the ``bluest expansion velocity'' (highest velocity at which we can measure a blue-shifted absorption) for $\mathrm{H\alpha}$, $\mathrm{H\beta}$, and Fe~II~$\lambda$5169.

\subsection{First-light epoch estimations}
Our initial sample was constructed by considering SNe~II with published non-detections on the Transient Name Server (TNS\footnote{\url{https://www.wis-tns.org/}}) occurring at most four days before the first detection. However, in some cases these published values (that are often automatically sent to the TNS) are not particularly constraining for the first-light  epoch, for example when the ``non-detection'' provides an upper limit that is actually brighter than the discovery photometry. Therefore, we re-analysed our full sample to robustly define the non-detection and first detection and thus the estimated first-light epoch. We only used ATLAS forced photometry values for non-detections and detections to maintain consistency and avoid any potential systematic effects from information from other surveys that have different 
sensitivity and cadence. This was the approach even when the discovery of a SN was first reported by another survey. Since ATLAS obtains four points per epoch bin, we calculated the weighted average of the flux measurements in each bin corresponding to the detection epoch ($\mathrm{MJD_{disc}}$) to obtain the detection flux. To determine the last deep non detection, we also calculated the weighted average of the flux measurements per epoch prior to the reported detection epoch. For both detection and non-detection we inferred the weighted standard deviation based on the individual measurement uncertainties. We considered the flux prior to detection to be deep enough for a non-detection if: 1) the forced photometry value was consistent with zero flux within one sigma, and 2) the error bars of the detection and non-detection did not overlap. When these two conditions were met, we adopted these as the last non-detection epoch ($\mathrm{MJD_{non-det}}$) and first detection epoch. An example of this method is shown in Figure \ref{fig:example_lc}. 

For approximately 10~\% of the sample, we found that the errors overlapped or that the flux at non-detection was not consistent with zero. In those cases, we used the previous forced photometry value before the original non-detection or the next flux value after the detection, depending on the situation, and iterated the process. As our sample was constructed to ensure a deep non-detection less than four days before the detection, we verified that this condition was still met. For four SNe, SN~2020uwl, SN~2021dbg, SN~2022cvu, and SN~2022fli, this was not satisfied. To maintain consistency, we removed these SNe from our sample. The first-light epochs for all the SNe were then calculated as the midpoint between the SN discovery date and the non-detection. The uncertainty associated with this epoch was taken as $(\mathrm{MJD_{disc}}-\mathrm{MJD_{non-det}})/2$. 

\subsection{Distance estimations}
\label{sec:distance}
The distances for the SNe~II are based on redshift when the recession velocity is larger than 2000~km~$\mathrm{s^{-1}}$. To calculate the luminosity distances, we used the following approximation:

\begin{equation}
    d_{L}=\dfrac{(1+z_{helio})}{(1+z_{CMB})}~\dfrac{c}{H_{0}}~\times~\Bigr[z_{CMB}~+~\dfrac{1}{2}\Bigl(\Omega_{\Lambda}-\dfrac{\Omega_{M}}{2}+1\Bigl)~+~z_{CMB}^2\Bigr],
\end{equation}

\noindent where $z_{helio}$ is the heliocentric redshift of the host galaxy,
$z_{CMB}$ is the redshift referred to the CMB rest frame, and $H_{0}$ and $\Omega_{M}$ are set to $H_{0} = 67.4 \pm 0.5$~km~s$^{-1}$~Mpc$^{-1}$ and $\Omega_{M} = 0.315 \pm 0.007$, respectively \citep{planck18}. The factor $\mathrm{(1+z_{helio})/(1+z_{CMB})}$ accounts for the fact that the photon redshift is observed with respect to the heliocentric reference system. 
The heliocentric redshift was calculated from the measurements of narrow emission lines originating from the host H II regions within the SN spectrum, then the $z_{CMB}$ was obtained from the NASA Extragalactic Database (NED\footnote{\url{https://ned.ipac.caltech.edu/}}) velocity correction calculator. We estimated the redshift error (and propagated this to our distance error) considering: 1) the standard deviation of our measurements, and 2) the spectral resolution. A typical $\sigma_{z_{helio}}=0.003$ was obtained. If we could not extract the H~II regions either because the spectrum was noisy or because the galaxy was faint, we used the host galaxy redshift from NED, and the error reported there. If a host galaxy had recession velocities less than 2000~km~s$^{-1}$, then a redshift-independent distance was taken from NED if available. This was the case for only one SN in the sample, SN~2021yja. For those SNe for which we could not resolve H~II regions and their host galaxies did not have NED information, we used a redshift value obtained from spectral matching with SNID, and employed the inferred distance. In those cases we took the fit error that comes from SNID, with typical values of $\sigma_{z}=0.003-0.01$.

Following the above procedure, there were 20 SNe for which we could estimate the redshift from H~II regions, 37 SNe where we got the redshift from NED, 12 SNe for which the redshift was estimated by spectral matching, and 1 SN for which we used a redshift-independent distance.
The mean redshift value of the sample is 0.023 and the median is 0.021.

\subsection{Extinction}
\label{sec:ext}

To correct for the line of sight extinction produced through absorption by intervening dust in the Milky Way, we take the E(B$-$V) from the \citet{schlafly11} recalibration of the \citet{schlegel98} infrared-based dust map, available from NED, and together with an extinction law from \citet{fitzpatrick99}, with $R_{V}=3.1$ we obtain $A_{V}$. To convert $A_{V}$ values to the ATLAS bands we used the Python module \textit{extinction}\footnote{\url{https://extinction.readthedocs.io/en/latest/}}.

We do not correct for host-galaxy extinction. Several methods exist in the literature to estimate host extinction. These include the Na I D equivalent width \citep{turatto03,rodriguez23}, the diffuse interstellar band (DIB) at 5780~{\AA} \citep{phillips13}, the (V$-$I) colour excess at the end of the plateau \citep{olivares10}, the colour-colour curve method \citep{rodriguez14,rodriguez19}, and the Balmer decrement \citep[e.g.][]{dominguez13}. However, there are major caveats with all of these methods. In this sample, we have low-resolution optical spectra for a subset of SNe, but the Na I absorption in low-resolution spectra has been shown to be a bad proxy for measuring the extinction \citep{poznanski11,phillips13}. Additionally, the DIB relation requires moderate resolution and high signal-to-noise ratio spectra \citep{phillips13}, but such observations are lacking in our sample. From optical spectra one could measure the Balmer decrement but, it is unclear whether the SN exploded behind, in front of, or within the H~II region. Concerning optical colours, our sample lacks extensive colour information. Even if we had more data, these methods assume that all SNe~II share the same intrinsic colour at some epoch. However, \citet{dejaeger18} argued against such a simplistic scenario in an in-depth study of SN~II colours. Furthermore, \citet{faran14} analyzed a sample of SNe~II and concluded that no extinction correction method significantly improved the uniformity of the sample. This finding was supported by \citet{gutierrez17b}, who discovered that stronger correlations (between various SN observables) within a SN~II sample were obtained before applying any extinction corrections. Therefore, we decided to not correct for host extinction since any such corrections are likely to simply add noise to our results. However, we note that this is a caveat in our work. 

\subsection{Photometric properties}
\subsubsection{ATLAS \texorpdfstring{$c$}{c}- and \texorpdfstring{$o$}{o}-band photometry}
After obtaining all ATLAS forced photometry for our sample as outlined in Section \ref{sec:sample}, we filtered the LCs by flux error, removing data with errors larger than 50 $\mu$Jy to remove spurious data points. Then, our analysis goal was to have robust rise-time estimations. To achieve this, we required sufficiently well-sampled LCs during the rise to maximum. Given the fast rise of SNe~II, we defined this criterion to be at least one photometric epoch between first light and maximum. While the LCs in the $o$-band are typically well-sampled, those in the $c$-band often lack photometric points during the rise. We calculated the photometric parameters (as outlined below) for the entire sample in the $o$-band, and for 21 events in the $c$-band, which were the only SNe that had data during the rising phase. Since ATLAS $o$-band light-curves are rather nosiy (photometric data points taken on the same night have large dispersion), after filtering by error, we still had substantial dispersion and a number of clear outliers. We thus decided to perform sigma-clipping on the LCs.  Given the intrinsic variability in the data around maximum, we did not sigma-clip the values during the early, rapidly rising phase of the LC, from the time of first light to maximum light, to prevent losing useful data in the process; clipping was performed at epochs later than 10 days after first light. We used this 10-day limit because it is in agreement with the mean rise times found in previous works (see for example  \citealt{gonzalezgaitan15}, \citealt{gall15}, \citealt{pessi19}). We median sigma-clipped the values using a 4$\sigma$ limit and a rolling window of 20 points. This is the range where the median distribution is calculated, which corresponds approximately to five days. $c$-band photometry displays less dispersion than the $o$-band photometry, so we did not perform sigma-clipping.

\subsubsection{ATLAS SNe~II light-curve measurements}
\label{sec:bazin}
We calculated the time of maximum light for our sample of SNe~II  by fitting the Bazin function \citep{bazin09} to the LCs. This is a phenomenological function that has the form:
\begin{equation}
    f(t)=A \frac{e^{-(t-\mathrm{t_{0}})/\mathrm{\tau_{fall}}}}{1+e^{(t-\mathrm{t_{0}})/\mathrm{\tau_{rise}}}} + B,
\end{equation}

\noindent where $\mathrm{\tau_{rise}}$, $\mathrm{\tau_{fall}}$, $\mathrm{t_{0}}$, A and B are treated as free parameters. $\mathrm{\tau_{rise}}$ and $\mathrm{\tau_{fall}}$ are related to the rise and fall of the LC,  A and B are normalization and scaling factors, respectively, and $t_{0}$ is the inflection point for the function. The date of maximum light ($t_{max}$) can be calculated by taking the derivative of this function, as explained in \citet{gonzalezgaitan15}. The fits were calculated using Markov chain Monte Carlo (MCMC) methods with the python emcee package \citep{foreman-mackey13}. This approach involves computing the posterior probability of the parameters given the observations, based on assumed prior distributions. The prior distributions were constructed by first performing a single fit for each SN using the python package SciPy to obtain initial best-fit parameters. Distributions for each parameter were then centered on these initial values.
We employed 100 walkers and 10000 steps per walker, together with a burn-in period of 2000 steps. Next, the date of maximum light was computed taking the derivative of the Bazin function. We obtained a $t_{max}$ value for each MCMC realization and then computed the mean and standard deviation. The rise times ($t_{rise_c}$ and $t_{rise_o}$) are then calculated as the difference between $t_{max}$ and the time of first light. The uncertainty in the rise time is the result of adding the uncertainty in the first-light calculation and the uncertainties in the time of maximum in quadrature. We note that some SNe show two peaks in their LCs, they are: SN~2021tyw, SN~2021aatd (87A-like), SN~2022acko, and SN~2022acrv (87A-like). In those cases, we only fitted the first peak with the Bazin function. 

Maximum-light absolute magnitudes ($M_{max_c}$ and $M_{max_o}$) were calculated using the distances and extinctions defined in Sections \ref{sec:distance} and \ref{sec:ext}, respectively, and the apparent magnitudes at $t_{max}$. The uncertainty in the absolute magnitude at maximum is the result of adding in quadrature the uncertainty in the maximum flux calculation and the error in the distance. 

For the decline rates, we adopted the definitions provided in \citet{anderson14}, where $s_{1}$ represents the decline rate of the initial, steeper slope of the LC, and $s_{2}$ corresponds to the decline rate of the second, shallower slope in the LC. It is important to note that while \citet{anderson14} express these values in magnitudes per 100 days, we calculate them as the change in flux normalized to maximum per 100 days. We call it $\Delta$f / 100 days. We first normalized the LCs to their maximum flux. Then, piecewise linear fits were applied to the photometry, testing both single- and two-slope models to assess whether the LCs exhibit different evolutionary phases or follow a single trend. The fitting range extended from the time of maximum light to either the point where the LC becomes noise-dominated (we visually inspect the LCs to determine when this happen) or the end of the plateau phase, whichever occurred first. Then we used the Bayesian information criteria (BIC) to determine which fit was better. If the BIC favored one slope, then we considered $s_{1}$=$s_{2}$, while if the BIC favored two slopes, we considered two different values for $s_{1}$ and $s_{2}$. The uncertainties on the decline rates were estimated from the fit error. The fits to the LCs are shown in Appendix~\ref{ap:2}. For those SNe with two peaks in their LCs, again we only considered the first peak.

\subsection{Spectral measurements}
\label{sec:spec_measurments}
\begin{figure}
	\includegraphics[width=0.97\columnwidth]{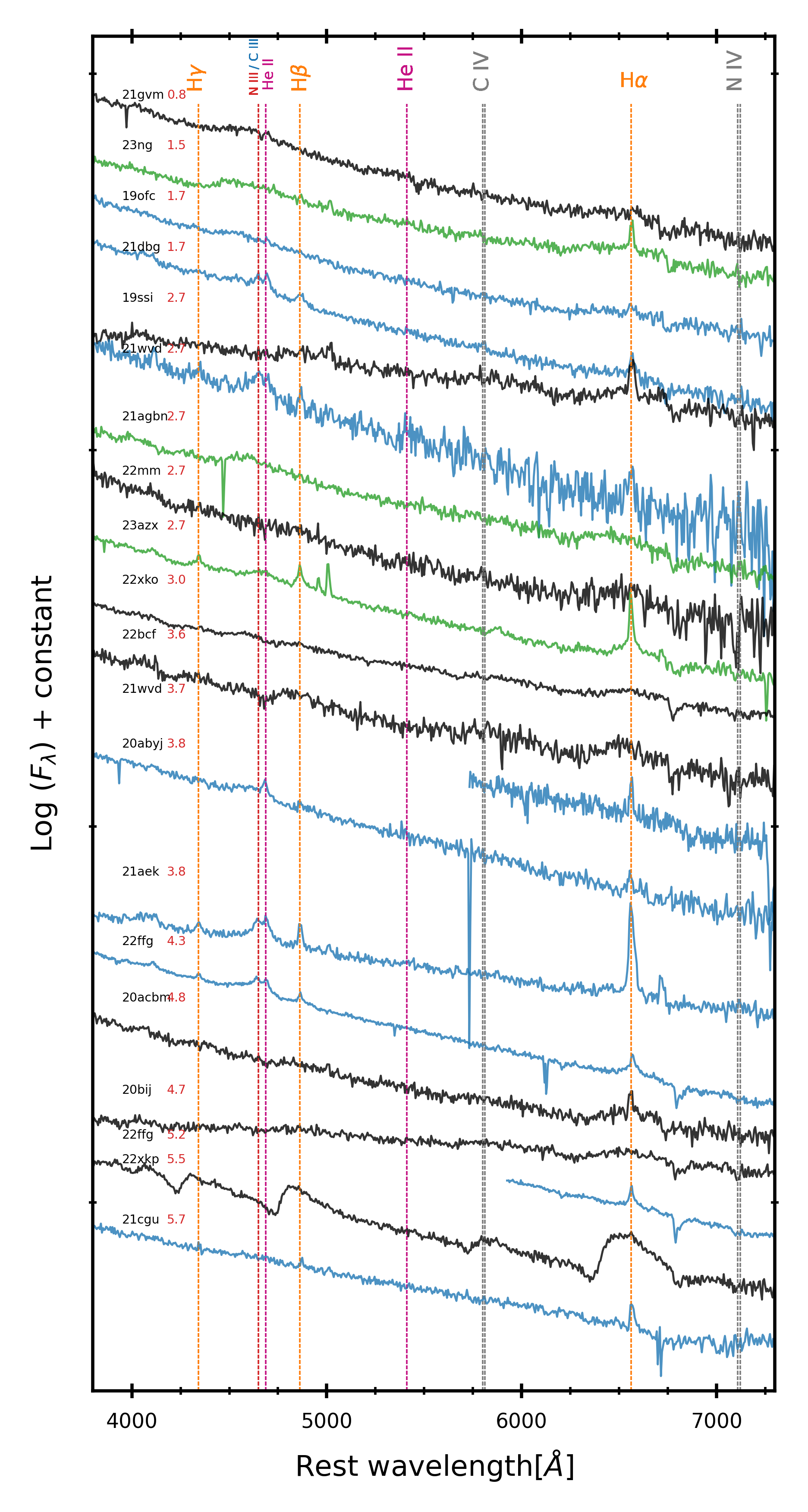}
    \caption{Spectra of the sample taken within 6 days past first light. Marked with vertical dashed lines are the main emission lines in the early phase. Lines in gray mark typical flash emission features that we do not observe in our spectra.  Non-flash SNe are marked with black, flash SNe  with blue, and ledgers with green (see Section~\ref{sec:class}). The colour scheme is kept for the rest of the paper.}
    \label{fig:spec_vearly}
\end{figure}

A  focus of our work is to study how any possibly-present CSM affecting the early photometric SN~II properties could also affect spectral properties observed at later times. In this work, we define early spectra as those obtained at $t$ < 25 days after first light. This limit at 25 days is roughly when the recombination of hydrogen becomes the principal physical mechanism responsible for the escape of the shock deposited energy \citep{bersten11,pumo11,pumo25}. We then refer to spectra taken at more than 25 days past first-light as photospheric spectra. There are 40 SNe in the sample with early spectra, 18 of which show a featureless continuum, while 22 have measurable spectral features (that will be described in Section \ref{sec:class}). Within the early spectra, there is a subset taken less than six days after first light, this subset is particularly interesting because it is in the range where the flash features are often produced \citep{bruch21,jacobsongalan24}. We show this subset in Figure \ref{fig:spec_vearly}, and we analyze it in Section \ref{sec:class}.

\begin{figure}
    \centering
    \subfloat{{\includegraphics[width=0.9\columnwidth]{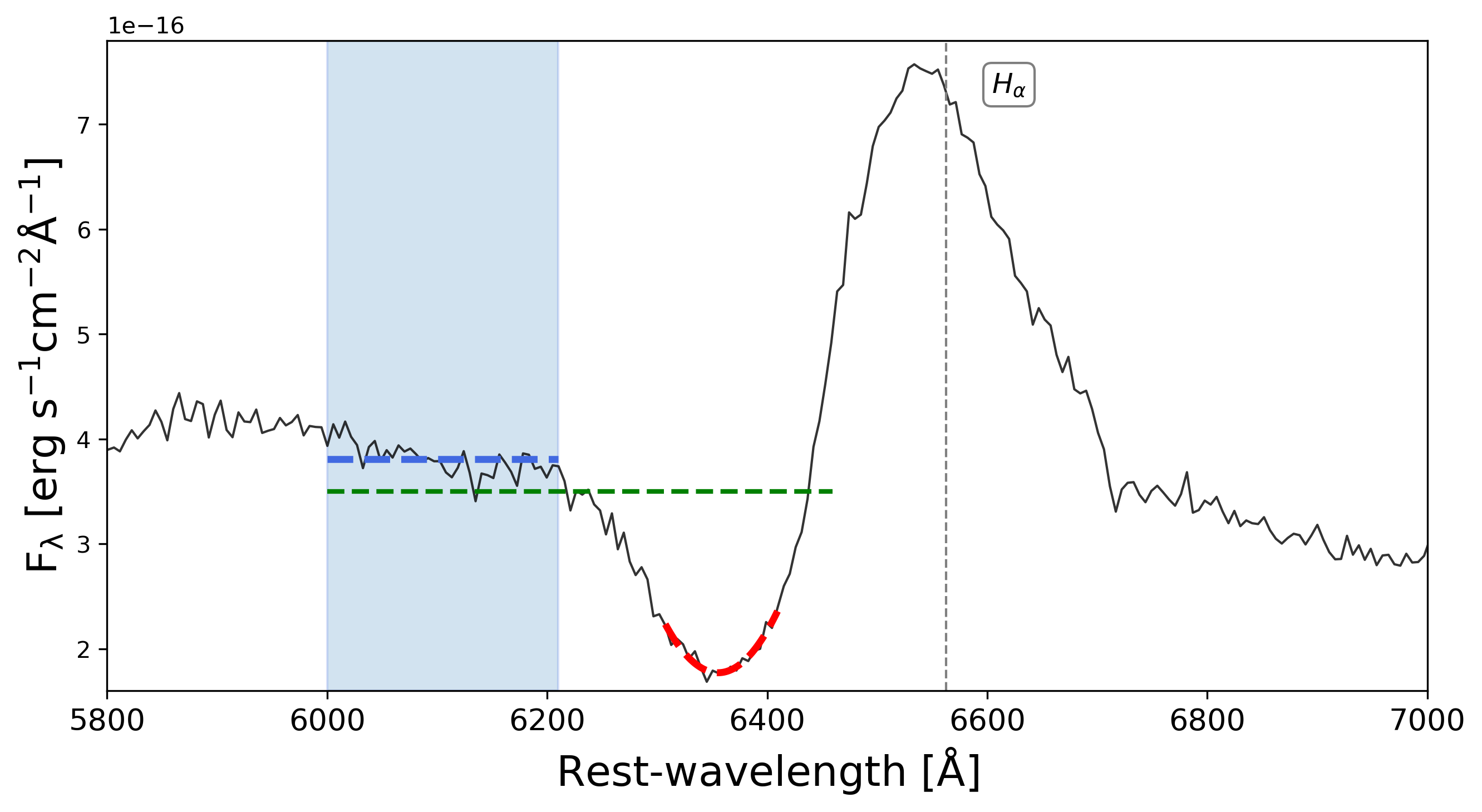} }}%
    \qquad
    \subfloat{{\includegraphics[width=0.9\columnwidth]{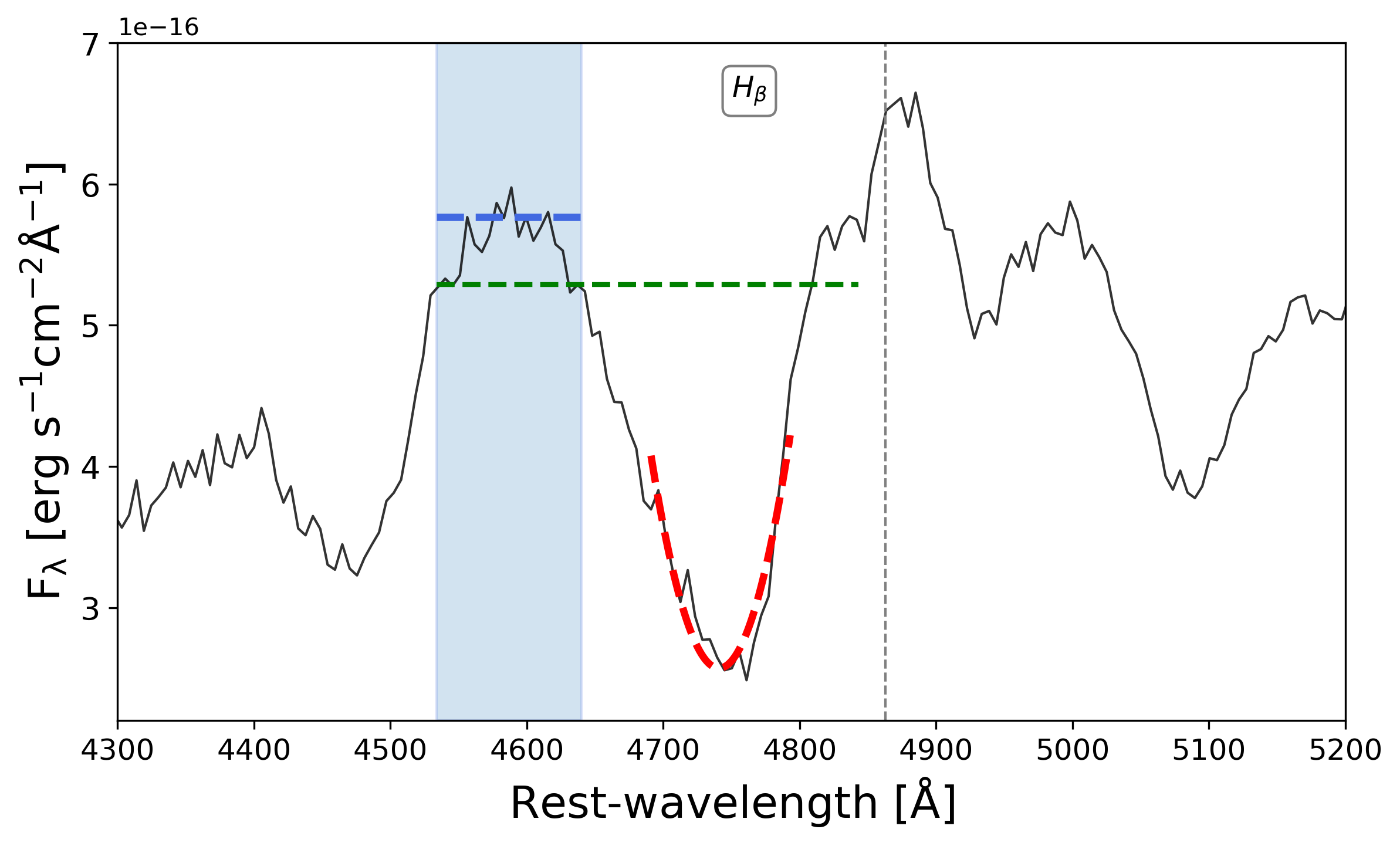} }}%
    \caption{Example of the fits and measurements of the minimum and the bluest wavelength for $\mathrm{H\alpha}$ (top) and $H_{\beta}$ (bottom). Marked in blue dashed lines are the fits to the pseudo-continuum to the blue side of $\mathrm{H\alpha}$ (top), and on the feature next to $H_{\beta}$ (bottom). The fits to the absorption minimum are marked in red for both panels, while the green line represents the value of the flux for which we consider the bluest wavelength of the feature.}
    \label{fig:fits_ha_hb}
\end{figure}

In our sample, we also have 45 SNe with photospheric spectra.
Since we are interested in comparing photospheric properties with spectral measurements, especially of the Balmer features, we create a sub-sample of SNe to achieve this purpose. We collected all SNe that have photospheric spectra, with the requirement that they also have a measurable $\mathrm{H\beta}$ profile. We define ``measurable'' when the $\mathrm{H\beta}$ absorption is strong enough to be able to fit a low order polynomial to the minimum. Twenty one SNe~II fulfill this criterion. For these SNe, we measured the expansion velocities of $\mathrm{H\alpha}$, $\mathrm{H\beta}$, and Fe~II~$\lambda$5169 by fitting a low order polynomial to the minimum of the absorption profiles. The absorption minimum is often considered a reliable indicator of the bulk velocity of the material producing that line. An example of the minimum fit can be seen in Figure \ref{fig:fits_ha_hb}.

We relate the fastest material of the ejecta with the bluest wavelength of an absorption feature. We are interested in this information because if significant CSM material exists in SNe~II, then it will dampen this fastest material as the kinetic energy is converted into luminosity. Our method to measure this bluest wavelength was to define a pseudo-continuum level blueward of the absorption feature and then estimate the bluest extent of the absorption with respect to this defined continuum. We thus defined the bluest wavelength as the wavelength where the flux is lower than the continuum level by 15\% of the maximum absorption. The 15\% level was decided as a compromise between attempting to measure as close to the continuum level as possible and avoiding the typical noise fluctuation of our spectra. For $\mathrm{H\alpha}$ measurements, we defined the level of the continuum by fitting a straight line in the range of 6000-6200~{\AA}. $\mathrm{H\beta}$ has other features to the blue and there is no smooth pseudo-continuum. Therefore, we took the emission to the left of the absorption, fitted a low order polynomial to that feature and took its maximum value as the level of the continuum. An example of these techniques is shown in Figure \ref{fig:fits_ha_hb}. Most of the uncertainty in the velocities was estimated by considering the standard deviation of repeated measurements when slightly changing the bluest and reddest wavelengths for the measurements.

Across our sample, we observe significant diversity in the $\mathrm{H\alpha}$ profiles, both in the shape and the strength of the emission and absorption components. \citet{gutierrez14} quantified this diversity by measuring the ratio of the equivalent widths (EWs) of the absorption to emission (a/e) components of $\mathrm{H\alpha}$. Following their approach, we calculated this ratio by determining the EW of each component individually.

\subsection{Classification of SNe~II with high ionization lines}
\label{sec:class}

As described in Section \ref{sec:spec_measurments}, we obtained spectra for 18 SNe within the first six days post-first light (see Figure~\ref{fig:spec_vearly}) and for five other SNe between six and ten days post-first light. Examining these early spectra, we observed that while some SNe appear featureless, others display narrow He~II~$\lambda$4686 with electron-scattering broadened wings emission lines. This feature has been used in the literature to classify SNe~II with ``flash ionization'' features \citep{khazov16,jacobsongalan24}, assumed to be caused by interaction between the ejecta and the CSM. One of the primary goals of this work is to determine the frequency of CSM interaction and its impact on LCs and spectra at late times, making the identification of SNe with early interaction crucial. To achieve this, we classified SNe as either showing high-ionization features (``flash SNe'') or lacking them (``non-flash SNe''). With this purpose, we searched the Weizmann Interactive Supernova Data Repository (WISeREP\footnote{\url{https://www.wiserep.org/}}) for public spectra of 
SNe~II, to augment our sample with the earliest spectra available. This search yielded 28 SNe with publicly available spectra.

\begin{figure}
	\includegraphics[width=\columnwidth]{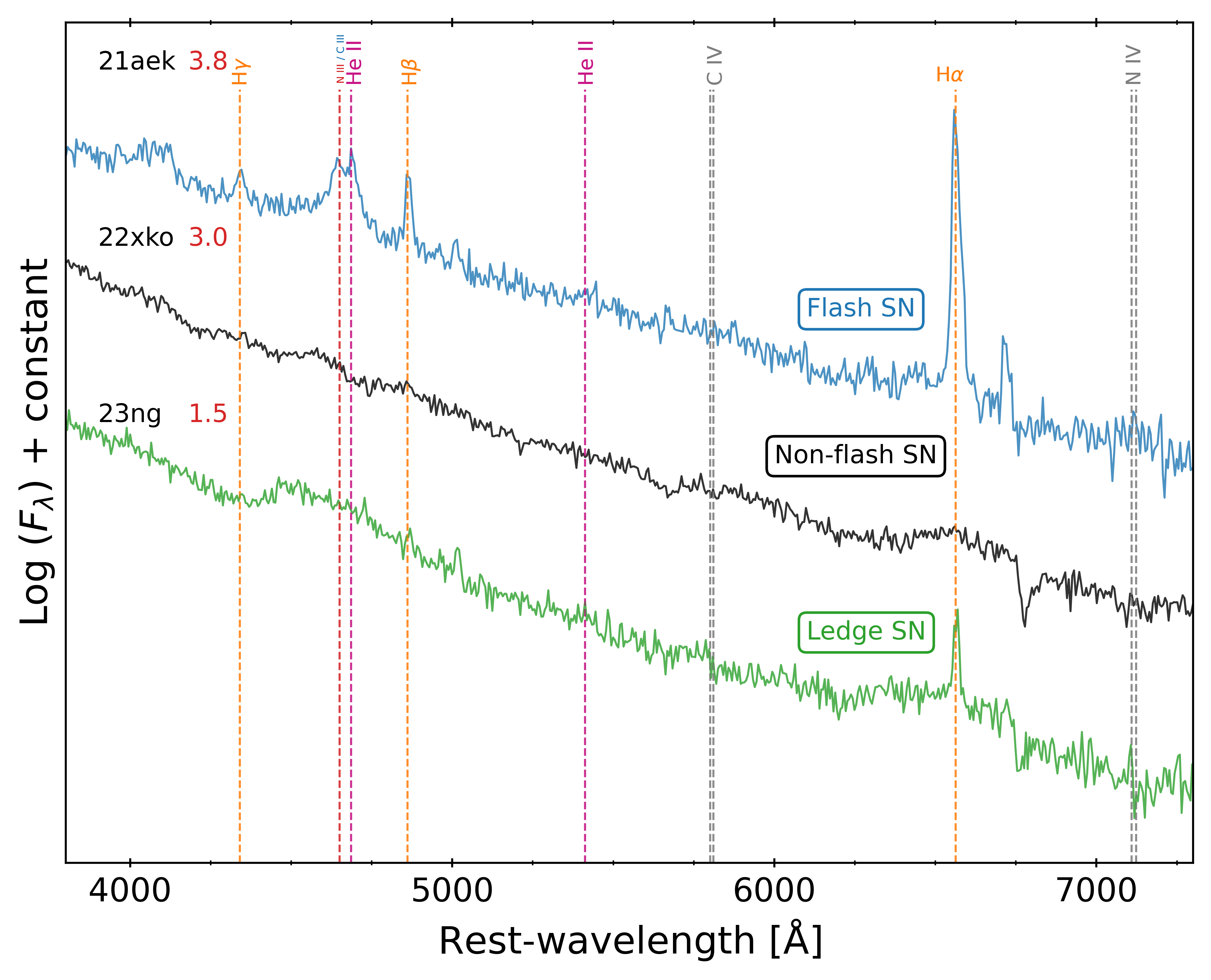}
    \caption{An example of a flash (blue), non-flash (black), and ledge (green) SN. Vertical lines mark the main emission lines during the early phase.}
    \label{fig:spec_class}
\end{figure}

Our sample includes 51 SNe with early spectra, combining our observations and public data. Of these, 42 SNe have spectra from 0 to 6 days after first light, while nine other have spectra from 6 to 10 days after first light. We classified them as flash or non-flash SNe by looking the profile of He~II~$\lambda$4686. Although this line is present in all SNe II spectra, when it is Doppler-broadened and blue-shifted it forms in fast moving material, but when it is narrow with electron-scattering broadened wings it forms in unshocked CSM. Only in the latter case this line is an unambiguous signature of interaction. Thus to classify the SNe as flash, we look for narrow He~II~$\lambda$4686 with electron-scattering broadened wings in its early spectra. To optimise objectivity, the authors KE, GF, and JA did this blindly, and discussed the results. The authors independently identified 18 SNe with high-ionization emission lines, including N~III and He~II. Additionally, we identified eight SNe displaying a broad emission feature around 4500–4800~{\AA} with an irregular profile, referred to in the literature as a ``ledge'' feature, which has also been claimed to be associated with interaction \citep{bullivant18,andrews19,soumagnac20}. These SNe were classified as ledge cases. The remaining 25 SNe showed no flash features and were categorized as non-flash. In total, our sample consists of 18 flash SNe, eight ledge SNe, and 25 non-flash SNe. The classification for each SN as well as the references for the public spectra are listed in Table~\ref{tab:sample2}. Additionally, an example of flash, non-flash, and ledge SNe is shown in Figure~\ref{fig:spec_class}.

The origin of the ledge feature remains ambiguous, as will be discussed in Sections \ref{sec:res} and \ref{sec:disc}. Given the ongoing debate about whether this feature indicates interaction, we consider the SNe classified as ledgers both as  SNe with signs of interaction and non-interaction in different analyses throughout the paper, to then compare our results under each assumption. 

We identify two primary sources of uncertainty in our classifications: 1) classifying SNe using spectra obtained after 6 days post-first light to search for flash features may not be reliable, and 2) some spectra (approximately 4) have low signal-to-noise, which could hinder our ability to determine if there is possible excess emission. To address these issues, we defined one sub-sample consisting of 38 SNe classified as flash or non-flash SNe with high signal-to-noise (above 15 according to \citealt{bruch21}) spectra taken within 0 to 6 days post-first light, and another sub-sample including the remaining 13 SNe classified with spectra taken from 6 to 10 days post-first light and low signal-to-noise.

\section{Results}
\label{sec:res}
\subsection{Light-curve property distributions}
\label{sec:lc_dist}
\begin{figure}
	\includegraphics[width=\columnwidth]{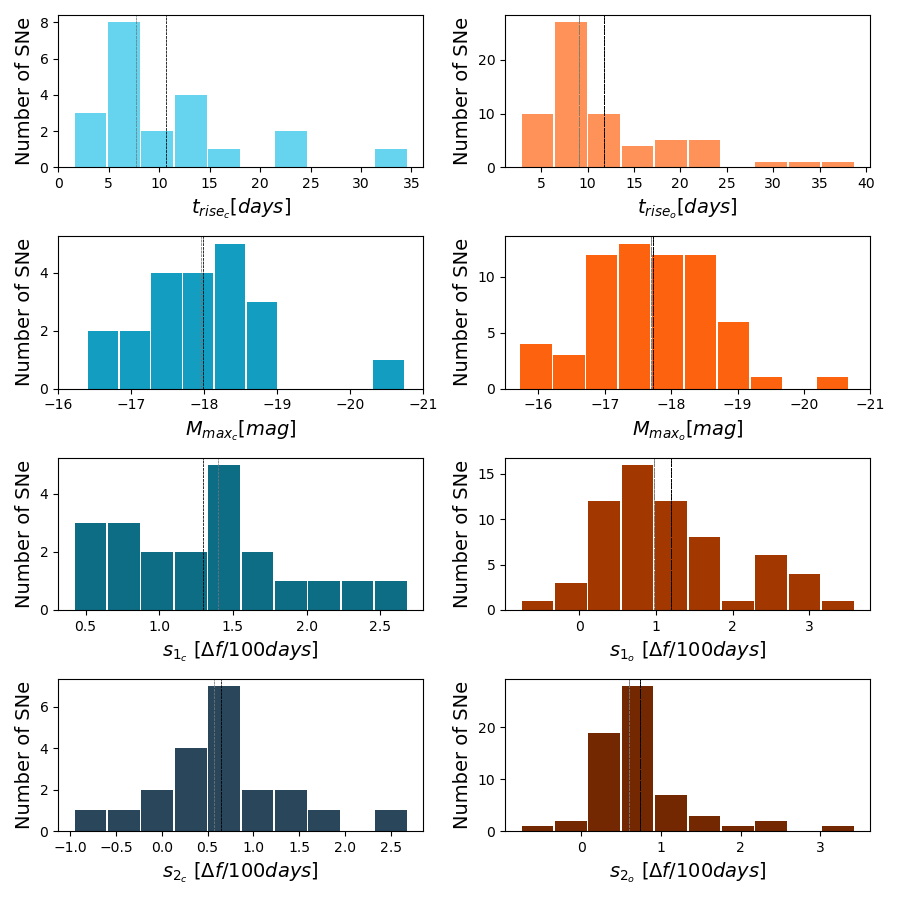}
    \caption{Distribution of rise times, absolute magnitudes at maximum light, and decline rates for the SNe in the sample. Vertical dashed lines in black and gray indicates the mean and median of each distribution, respectively. This figure does not contain the 87A-like events. Left panel: ATLAS $c$-band. Right panel: ATLAS $o$-band}
    \label{fig:stats}
\end{figure}

The distributions of rise times, absolute magnitudes and decline rates for both ATLAS bands are shown in Figure \ref{fig:stats}. SN~1987A-like events are not included in this figure (nor subsequent ones), nor in presented distributions means, medians, etc.

The mean rise time of the sample in the $o$ band is $11.8$ days ($\sigma=7.1$~days, 64 SNe), while the median is $9.1$ days; in the $c$ band the mean rise time is $10.7$ ($\sigma=7.7$~days, 21 SNe) and the median $7.7$ days. The two longest rise times correspond to SN~2020vef and SN~2022iek, with values in the $o$ band of $38.8$ and $31.8$ days, respectively; while the shortest rise times correspond to SN~2021sqg and SN~2020abyj with $o$ band values of $1.9$ and $2.2$ days, respectively. The four 87A-like events, SN~2021aatd, SN~2021adcw, SN~2021adyl, and SN~2022acrv have rise times of $86.3$, $52.9$, $56.2$, and $24.8$ days, respectively. Note that SN~2021aatd and SN~2022acrv present additional early peaks in the LC, with initial rise times of $12.8$ and $3.2$ days, respectively.

Although there is no calculation of rise times in the literature for the ATLAS bands, in Table \ref{tab:LCdist} we compare our results with those of previous works that employed defferent bands. The $o$ filter is roughly equivalent to $r+i$ filters, so it is suitable to compare our results with those previously calculated in the $R$/$r$ bands. Our results are in agreement with those of \citet{gonzalezgaitan15} for wavelengths > 7000~{\AA}, but are somewhat higher than those calculated by \citet{gall15} and \citet{rubin16} and smaller than that calculated by \citet{pessi19}. Although the dispersion in the literature values is high, our results are in agreement overall. The differences could be due to the fact that we are using a very broad filter. 
Note that, similar to \citet{gonzalezgaitan15}, we identify a subset of the sample with long rise times (i.e. longer than 15 d). We will discuss this issue in detail in Section \ref{sec:disc_rise}.
 
Focusing on the absolute magnitude distributions, the mean is $-17.7$ mag ($\sigma=0.9$~mag, 64 SNe) and the median is $-17.7$ mag in the $o$ band; in the $c$ band the mean is $-18.0$ mag ($\sigma=0.9$~mag, 21 SNe) and the median is $-18.0$ mag. In Table~\ref{tab:LCdist} we compare our absolute peak magnitudes with those previously presented in the literature. There is no study to date that made these estimations for the ATLAS bands, so a direct comparison is not possible. However, the absolute magnitudes at maximum light we inferred are consistent within the dispersion, although slightly brighter than previous values available in the literature. 
The most luminous events in our sample are SN~2020znl and SN~2022yjl, while the dimmest are SN~2022xkp, SN~2022acko, and SN~2022acug. The bright events should be taken with caution because their distances may be not reliable. SN~2020znl exploded in the galaxy PS1-88920933245169840, which has no reported distance estimation. We also could not extract the H~II region from its spectra (to enable an accurate redshift estimate), so we estimated its redshift by spectral matching with SNID, as stated in Section \ref{sec:distance}. Although its distance is not reliable, SN~2020znl does present a spectrum typical of luminous SNe~II, with lack of $\mathrm{H\alpha}$ absorption \citep{pessi23}. SN~2022yjl is an even more difficult case since it does not have an identified galaxy, and we could not get any spectrum to extract the H~II region. We used the photo redshift value from its classification report \citep{li23}. We lack a photospheric spectrum of SN 2022yjl to search for further evidence of high luminosity.

The mean $s_{1}$ is $1.2$~$\Delta$f/100~days ($\sigma=0.9$~$\Delta$f/100~days, 64 SNe) in the $o$ band and $1.3$~$\Delta$f/100~days ($\sigma=0.6$~$\Delta$f/100~days, 21 SNe) in the $c$ band. For $s_{2}$, the mean is $0.7$~$\Delta$f/100~days for both the $o$ band ($\sigma=0.6$~$\Delta$f/100~days, 64 SNe) and the $c$ band ($\sigma=0.7$~$\Delta$f/100~days, 21 SNe). There are some objects in our sample that have large decline rates. These are SN~2020znl, SN~2023cr, SN~2023dr, and SN~2023azx. It is worth noting all these SNe are either flash or ledgers. We will discuss this further in Section \ref{sec:corr}.
We do not compare our decline rates with previous studies because, to maintain consistency throughout the paper, we express them in $\Delta$ flux per 100 days rather than magnitudes, preventing direct comparisons. 

Two of the four SNe that show double-peaked LCs are 1987A-like events, so they were not included in the distributions, while the other two SNe were included. Although these last two may have a different behaviour compared to the rest of the sample, we leave them in the distributions since they are confirmed type II SNe. However, we note that all their parameters are calculated with respect to their first peak.

\subsection{Spectral distributions}
\label{sec:spec_dist}
\begin{figure}
	\includegraphics[width=\columnwidth]{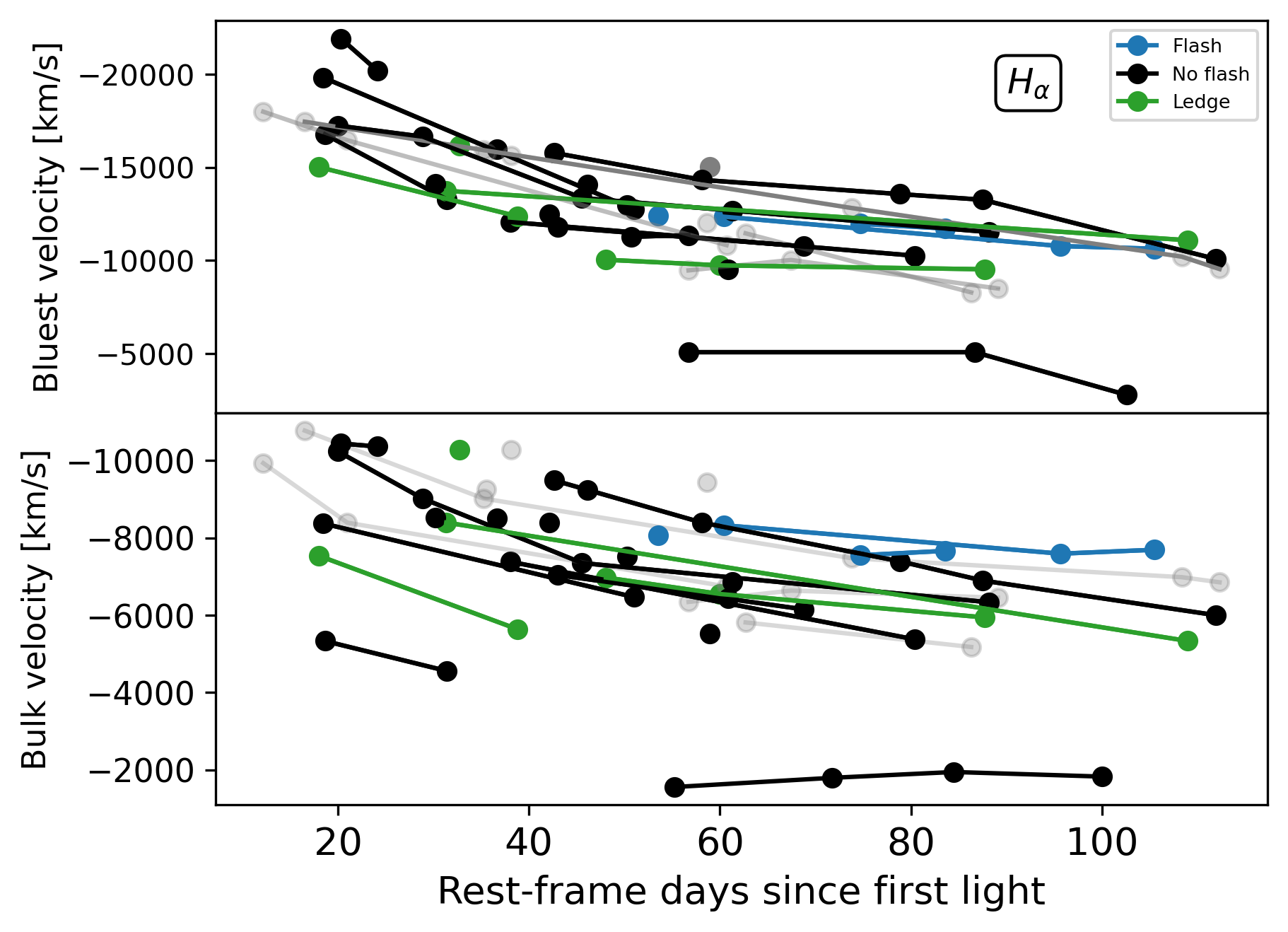}
    \caption{$\mathrm{H\alpha}$ bulk (upper panel) and bluest (lower panel) velocity evolution. Non-flash SNe are marked with black, flash SNe  with blue, ledgers with green,  and SNe without early spectra for flash classification are marked with gray. The errors are plotted but generally fall inside the data points.}
    \label{fig:vel_ev}
\end{figure}

We focus our study of SN~II spectral features by investigating the properties of $\mathrm{H\alpha}$, $\mathrm{H\beta}$, and Fe~II~$\lambda$5169. As an example, we show the $\mathrm{H\alpha}$ bulk and bluest velocity evolutions in Figure \ref{fig:vel_ev}. 

$\mathrm{H\alpha}$ shows a wide range of bulk velocities: from 5000 to 11000 km s$^{-1}$ during early phases and from 6000 to 9500 km s$^{-1}$ for most SNe in the photospheric phase. This is similar to what was obtained by \citet{gutierrez17b}, with $\mathrm{H\alpha}$ expansion velocities ranging from 9500 to 1500 km s$^{-1}$ at 50 days. For both $\mathrm{H\alpha}$ and $\mathrm{H\beta}$ the lowest values of bulk velocities are those of SN~2020aze and SN~2020vef. For $\mathrm{H\beta}$ also SN~2022acmt is in the range of low values, but it is not present in the $\mathrm{H\alpha}$ plot since it does not present an absorption component of $\mathrm{H\alpha}$. We could not measure the Fe~II~$\lambda$5169 velocity for SN~2020aze and SN~2020vef since the feature was very weak.  

When looking at the bluest velocities, SN~2020aze and SN~2022acmt appear well within the the rest of the distribution, while SN~2020vef remains with an extreme value. Other properties of these three SNe are: SN~2020aze shows the "Cachito" component in $\mathrm{H\alpha}$, also it is a part of the SNe with long rise times. SN~2020vef has a very weak $\mathrm{H\alpha}$ absorption, and it also has a long rise time. 

\subsection{Correlations}
\label{sec:corr}
\begin{figure*}
\centering
	\includegraphics[scale=0.69]{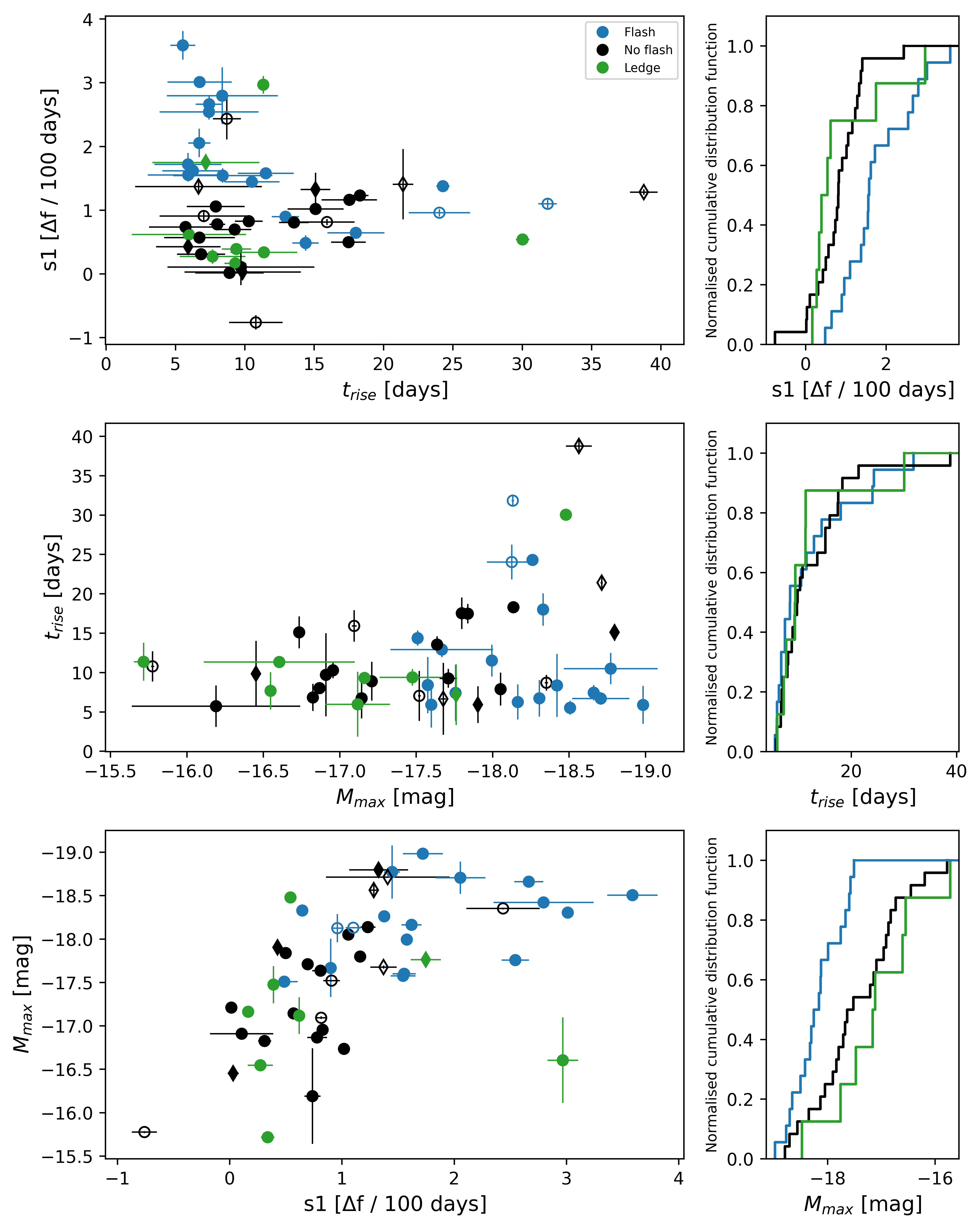}
    \caption{Correlations between photometric parameters in the sample. Non-flash SNe are marked with black, while flash SNe are marked with blue. Ledgers are marked with green and SNe without early spectra for flash classification are marked with gray. Filled circles indicate SNe classified as flash or non-flash based on spectra obtained within six days of first light and with high signal-to-noise. Diamond symbols represent classifications based on low signal-to-noise spectra, while open symbols correspond to SNe classified with spectra taken between 6 and 10 days after first light.
    }
    \label{fig:corr_phot}
\end{figure*}

Correlations between photometric parameters in the $o$-band, distinguishing flash SNe, non-flash SNe, and ledgers, are presented in Figure \ref{fig:corr_phot}, as well as the cumulative distribution functions of the parameters. Table \ref{tab:phot_param_flash} provides the mean values for all photometric parameters for flash and non-flash SNe, with ledgers considered as part of each group separately. The table includes results for both the clean and unclean samples. Additionally, Table~\ref{tab:phot_KS} shows the results of Anderson-Darling (A-D) tests for different groups across all parameters. 

Flash and non-flash SNe exhibit significantly different mean $s_{1}$ decline rates, as illustrated in the upper panel of Figure \ref{fig:corr_phot}. The ledger SNe appear to be spread across both groups. As seen in Table~\ref{tab:phot_KS}, an A-D test reveals a statistically significant difference in $s_{1}$ between the flash and non flash SNe. When including SNe classified as flash or non-flash based on spectra obtained 6–10 days after first light and low S/N ratios, the p-value remains statistically significant. These results suggest that SNe exhibiting high-ionization lines in their early spectra decline faster than those without signs of interaction. If the ledger SNe are considered as part of the flash SNe, the difference between the two populations becomes less pronounced but remains significant. However, it stays almost the same when treating ledgers as non-flash SNe. This suggests that ledgers present a distribution of $s_{1}$ decline rates that is more similar to non-flash SNe than to flash SNe. On the other hand, we do not find different mean $s_{2}$ decline rates between flash and non-flash SNe.

Flash SNe are also systematically brighter than non-flash SNe in terms of $M_{max}$. For the SNe II sample with spectra within 6 days of first light, a A-D test returns a p-value of 0.001, demonstrating a significant difference. This trend persists when extending the analysis to include the SNe II with spectra within 10 days of first light. The difference also remains when ledger SNe are added, except in the case where they are considered as flash SNe in the 10-day sample, which weakens the distinction.

In contrast to decline rates and $M_{max}$, no significant difference is found in $t_{rise}$ between flash and non-flash SNe. A–D tests for both the 6-day and 10-day samples consistently yield p-values greater than 0.2, suggesting no evidence for distinct populations based on this parameter.

\begin{table}
\centering
	\caption{Mean values of the photometric parameters for the sub-samples of flash and non-flash SNe. SNe classified as ledgers are included in both distributions separately. N is the number of SNe in each case.}
	\label{tab:phot_param_flash}
	\begin{threeparttable}
    \resizebox{\columnwidth}{!}{
	\begin{tabular}{lccccc} 
	    \hline
	    \multicolumn{6}{|c|}{SN II sample with spectra within 6 days of first light} \\
		\hline
		   Classification & $M_{max}$ & $t_{rise}$ & $s_{1}$ & $s_{2}$ & N  \\
		\hline
		Only flash SNe   & -18.2 $\pm$ 0.5  & 10.0 $\pm$ 5.0  & 1.8 $\pm$ 0.8 & 0.7 $\pm$ 0.4 & 16\\
		Only non-flash SNe & -17.3 $\pm$ 0.6 & 11.1 $\pm$ 4.2  & 0.7 $\pm$ 0.3  &  0.5 $\pm$ 0.3 & 14 \\
		  $\mathrm{Flash \cup ledgers}$& -17.8 $\pm$ 0.8 & 10.7 $\pm$ 6.0& 1.5 $\pm$ 1.0 & 0.6 $\pm$ 0.4& 23\\
		$\mathrm{Non-flash \cup ledgers}$ & -17.2 $\pm$ 0.7  & 11.4 $\pm$ 5.6 & 0.7 $\pm$ 0.6 & 0.4 $\pm$ 0.3& 21\\ 
    	\hline
         \multicolumn{6}{|c|}{SN II sample with spectra within 10 days of first light} \\
    	\hline
    	Only flash SNe & -18.2 $\pm$ 0.4 & 12.0 $\pm$ 7.5 & 1.7 $\pm$ 0.8 & 0.7 $\pm$ 0.4 & 18\\
		Only non-flash SNe & -17.4 $\pm$ 0.7 & 12.6 $\pm$ 7.4 & 0.8 $\pm$ 0.6 & 0.8 $\pm$ 0.8 & 21\\
		$\mathrm{Flash \cup ledgers}$ & -17.6 $\pm$ 0.8 & 11.9 $\pm$ 7.4 & 1.5 $\pm$ 1.0  & 0.6 $\pm$ 0.4  & 26\\
		$\mathrm{Non-flash \cup ledgers}$& -17.3 $\pm$ 0.8 & 12.3 $\pm$ 7.3 & 0.8 $\pm$ 0.7 & 0.6 $\pm$ 0.7 & 29\\ 
		\hline
	\end{tabular}
    }
	\end{threeparttable}
\end{table}

\begin{table*}
\centering
	\caption{Results of the A-D tests for different groups. For each parameter and each group the $A^2$ statistic is listed, along with the p-value. Highlighted in bold are those p-values $<$ 0.05.}
	\label{tab:phot_KS}
	\begin{threeparttable}
	\hspace*{-4cm}\begin{tabular}{lcccc} 
	    \hline
	    \multicolumn{5}{|c|}{SN II sample with spectra within 6 days of first light} \\
		\hline
		   Classification & $M_{max}$ & $t_{rise}$ & $s_{1}$ & $s_{2}$  \\
		\hline
            Flash vs Non-flash SNe   & 7.601/\textbf{0.001} & -0.178/0.250  & 8.846/\textbf{0.001} & 0.863/0.145 \\
            $\mathrm{Flash \cup ledgers}$ vs Non-flash SNe  & 2.855/\textbf{0.02} &  -0.480/0.250 & 3.703/\textbf{0.001} &  -0.416/0.25  \\
            Flash SNe vs $\mathrm{Non-flash \cup  ledgers}$ & 9.843/\textbf{0.001} & 0.086/0.250 &  10.084/\textbf{0.001} & 2.859/\textbf{0.022}\\ 
    	\hline
         \multicolumn{5}{|c|}{SN II sample with spectra within 10 days of first light} \\
    	\hline
            Flash SNe vs Non-flash SNe   & 8.786/\textbf{0.001} & -0.374/0.250 & 8.227/\textbf{0.001} & 0.148/0.250 \\
		    $\mathrm{Flash \cup ledgers}$ vs Non-flash SNe  & 1.554/0.074 & -0.577/0.250 & 3.708/\textbf{0.01} & -0.190/0.250   \\
            Flash SNe vs $\mathrm{Non-flash \cup  ledgers}$ & 8.035/\textbf{0.001} & -0.341/0.250 & 8.262/\textbf{0.001} & 1.305/0.094\\ 
		\hline
	\end{tabular}
	\end{threeparttable}
\end{table*}

We performed a Pearson test for the $t_{rise}$ vs $s_{1}$, $t_{rise}$ vs $M_{max}$, and $M_{max}$ vs $s_{1}$ and found correlation coefficients of -0.22, -0.14, and -0.56, respectively. Among these, the only strong correlation identified was between $M_{max}$ and $s_{1}$. No significant correlation was found between $t_{rise}$ vs $M_{max}$ or $t_{rise}$ against the $s_{1}$ decline rate (see Section \ref{sec:disc_rise}).

Figure \ref{fig:a_e_ev} shows the evolution of the a/e ratio. SNe with the smallest a/e ratios correspond to those displaying high-ionization lines in their early-time spectra due to CSM interaction. Once again, the ledgers SNe appear to be grouped with the non-flash population. 

We studied the $\mathrm{H\alpha}$, $\mathrm{H\beta}$, and Fe~II~$\lambda$5169 bulk and bluest velocities against absolute magnitude, decline rates, and rise times, respectively. In some SNe, the $\mathrm{H\beta}$ feature was present and measurable, whereas $\mathrm{H\alpha}$ was either absent or had an unmeasurable bluest wavelength. We calculated the Pearson correlation coefficients for the velocities against all the photometric parameters and found moderate correlations for the known relation of bulk velocity of  $\mathrm{H\alpha}$, $\mathrm{H\beta}$, and Fe~II~$\lambda$5169 vs absolute magnitude. Among these, Fe~II~$\lambda$5169 exhibited the strongest and clearest correlation. No correlations were observed with the other photometric parameters.

\section{Discussion}
\label{sec:disc}
\subsection{The rise times of SNe~II}
\label{sec:disc_rise}
For an isolated star with no interaction with CSM, standard stellar evolution theory predicts that the rise to maximum light is primarily influenced by the radius of the progenitor star \citep{swartz91, gall15}. Larger progenitor radii result in SNe with longer rise times. Metallicity, indirectly, also plays a role, as it affects the pre-SN structure of the progenitor star \citep{chieffi03, tominaga11}. Specifically, lower metallicity leads to smaller radii, thereby impacting the rise times. However, \citet{gonzalezgaitan15} found typical rise times of less than 10 days. When compared with analytical models, they found progenitor radii that were, on average, significantly smaller than those measured for observed RSG, suggesting that the rise times are too short to be explained by an isolated star with no CSM. Supporting this finding, \citet{forster18} reported a steep rise lasting just a few days in their sample of SNe~II. By comparing these observations to models, they inferred high CSM densities for most events and concluded that this likely reflects the systematic detection of shock breakouts within a dense CSM. 

The relatively short rise times motivated modelling efforts of SNe II with CSM interaction. \citet{morozova17} first modelled three well-studied Type II events, finding that these were well-reproduced by RSG surrounded by dense CSM. The extent of the CSM and the wind velocity suggested that these stars likely experienced enhanced mass-loss activity during the last months to years of their lives. Then, \citet{moriya23} demonstrated that the rise times of SNe~II are predominantly influenced by the properties of the CSM, such as its radius, mass, and mass-loss rate, as well as the explosion energy. According to \citet{moriya18}, when the mass loss rate is sufficiently high ($\dot{M}$ > $10^{-3}$~$M_{\odot}$~$yr^{-1}$), the rise times are primarily governed by the mass-loss rates. Interestingly, the rise times initially decrease as $\dot{M}$ increases, but beyond a certain threshold, they begin to lengthen with further increases in $\dot{M}$.

Our results are broadly consistent with these previous studies, since the majority of SNe II in our sample exhibit short rise times. While the mean rise time in the $o$-band for our sample is 11.7 days, there is a fraction of SNe -- without taking into account 87A-like events -- with longer rise times (longer than 15 days\footnote{The choice of this value is arbitrary and it is based on the number adopted by \citet{gonzalezgaitan15} to separate SNe with short and long rise times.}). These are 16 SNe, representing $\approx$24$\%$ of the sample. Figure \ref{fig:corr_phot} shows that SNe with long rise times seem to have slow declines and also tend to be brighter. Regarding their velocities, it has been proposed that SNe II with long rise times and high $\mathrm{H\alpha}$ velocities may be a link between type IIs and IIbs \citep{davis13}. Although we do not have spectra for all the SNe with long rise times, those SNe that have show $\mathrm{H\alpha}$ velocities consistent with other normal type II SN samples \citep{gutierrez17}.

Previously, \citet{gonzalezgaitan15} identified a fraction of SNe in their sample with long rise times. Although classified as Type IIs based on photometry, these events lacked spectroscopic confirmation. Those authors suggested that these SNe might originate from a different progenitor system and argued that this provides strong evidence that typical SNe II are limited to short rise times, while other possible hydrogen-rich SNe may exhibit longer rise times. In contrast, all SNe with long rise times in our sample are spectroscopically confirmed as normal Type II. 

When taking into account the early classification as flash or non flash SNe, we find that 4 of the SNe with long rise times are flash SNe, 8 are non-flash SNe, 1 is a ledger and for three of them, we do not have early spectra to classify. Although modelling the progenitor properties of these SNe is beyond the scope of this paper, we can speculate on possible scenarios. For SNe with long rise times and no signs of CSM interaction, it is plausible that they originate from standard RSGs without surrounding CSM, consistent with predictions from standard stellar evolution theory. On the other hand, for SNe displaying early spectral evidence of interaction, the findings of \citet{moriya18} may provide a natural explanation in which sufficiently high, but not extreme, mass-loss rates can lead to extended rise times without producing SNe powered by interaction, as seen in SNe IIn.

Overall, our rise time distributions show that SNe with early spectral signatures of CSM interaction are not confined to either the short or long ends of the rise times range but are rather spread across the distribution. This could be due, as commented before, to the multiple factors that affect the rise times.

\subsection{Links between early-time and late time SN~II properties}
We find that flash SNe exhibit faster decline rates in $s_{1}$ and are brighter than those without signs of interaction. This result contrasts with previous studies, such as that of \citet{bruch23}, who reported no significant differences in brightness, colour, or rise times between SNe II with and without flash ionization features. In addition, \citet{bruch23} found no differences in magnitudes, redshift, or the epoch of the first spectrum between flash and non-flash SNe, indicating that selection effects did not influence their results. We examined these same properties in our sample and likewise found no significant differences. The discrepancy between the two studies may therefore arise from differences in sample size (only 17 events in \citealt{bruch23} work), or in the treatment of SNe showing the ledge or Doppler-broadened He~II profiles, which they treat as flash SNe.

Despite this tension, recent studies report trends consistent with our findings. For example, \citet{jacobsongalan25} found that flash SNe have significantly higher luminosities and decline rates at +50 days than non-flash SNe, while \citet{jiang25} reported higher post-peak decline rates for SNe with larger CSM masses. Together, these results support our conclusion that CSM properties influence not only the early-time spectra but also the overall LC evolution. More specifically, the differences between flash and non-flash SNe~II with respect to decline rates appear to be confined to the early $s_{1}$ decline - i.e. that from peak to the start of the plateau. This suggests that the presence and diversity of this initial decline rate may be dominated by the extent and profile of CSM above the progenitor density profile.

\begin{figure}
\centering
	\includegraphics[width=\columnwidth]{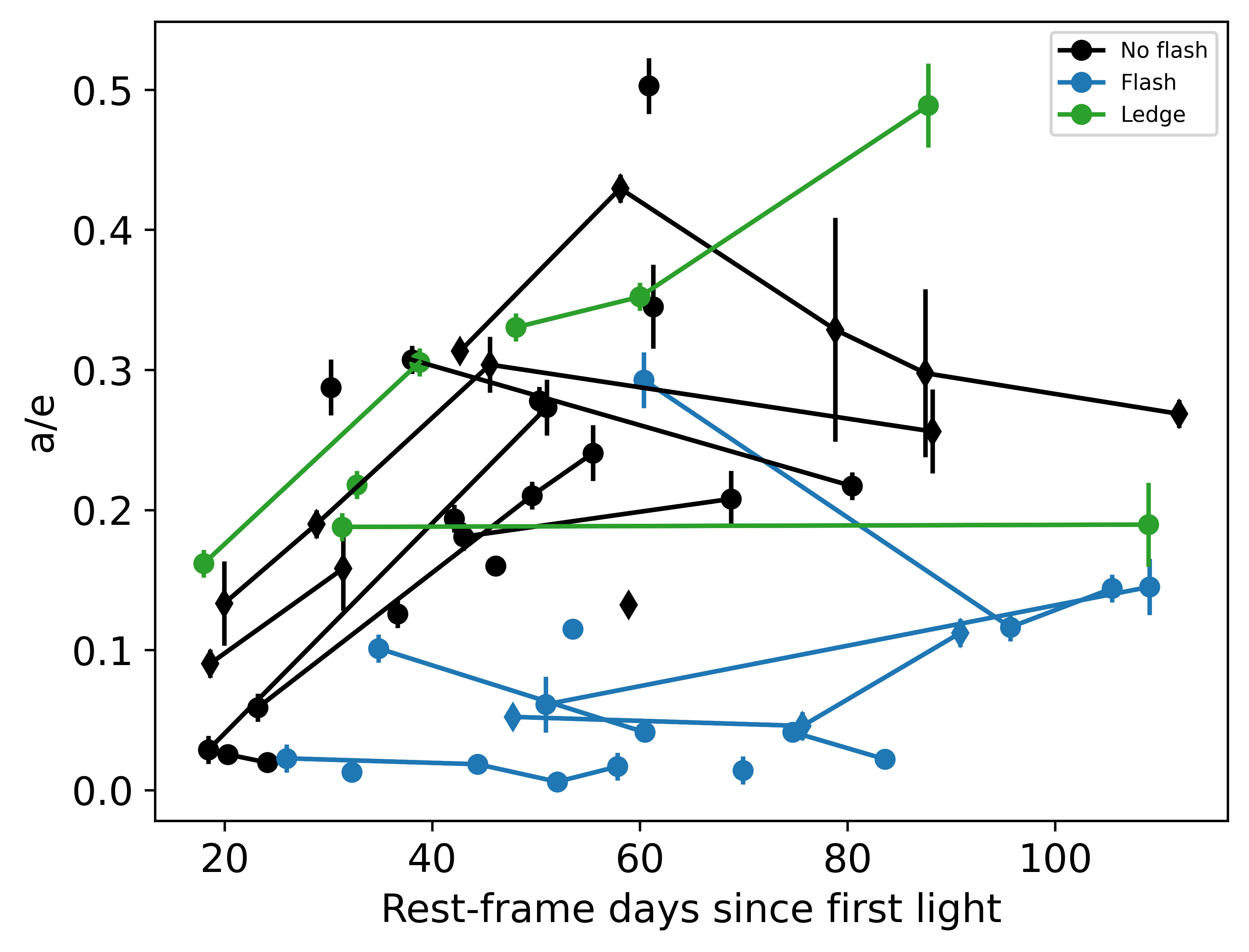}
    \caption{Evolution of the a/e ratio. Non-flash SNe are marked with black, flash SNe  with blue, and ledgers with green.}
    \label{fig:a_e_ev}
\end{figure}

An interesting finding of this work is the evolution of the a/e ratio, as shown in Figure \ref{fig:a_e_ev}. Our analysis reveals a distinct division: SNe that exhibit high-ionization lines in their early spectra tend to have smaller a/e ratios compared to those without such features. This ratio, which accounts for the diversity in the $\mathrm{H\alpha}$ absorption profile, has been previously studied by several authors. To explain the absence or weakening of the absorption component, \citet{schlegel96} proposed three plausible scenarios. First, the absorption component could be filled in by additional emission produced via scattering, potentially originating from CSM or a highly extended SN envelope. Second, it could be a low mass envelope which would have less matter to absorb, and no P-Cygni profile will be formed. Lastly, a steep density gradient in the hydrogen envelope could reduce the amount of absorbing material at high velocities, leading to the suppression of P-Cygni absorption. Additionally, \citet{gutierrez14} analysed the a/e ratio in a large sample of SNe~II and concluded that variations in the $\mathrm{H\alpha}$ P-Cygni profile are most likely associated with changes in the envelope properties (such as its mass and density profile) and the degree of CSM interaction. Furthermore, \citet{gutierrez14} found that SNe with smaller a/e have more rapidly declining LCs and are brighter at maximum light. Consistent with these findings, our results show that SNe with signs of interaction in the early spectra tend to be brighter, have rapidly declining LCs, and have the lowest a/e ratios, indicating a lack or depletion of $\mathrm{H\alpha}$ absorption. This aligns with the predictions of \citet{schlegel96} and \citet{gutierrez14}, reinforcing the role of CSM interaction in shaping the observed profiles.  

One matter of debate in this work is the association of ledge SNe with the other groups, not only in the a/e ratio evolution but also within the distribution of parameters. They appear to group with the non-flash population. Ledge features have been observed in early spectra several times in the literature \citep[e.g.][]{quimby07,shivvers15,bullivant18}, either alone or blended with a narrow component of high-ionization lines. However, its origin remains a subject of debate. One explanation is that the ledge arises from a very broad, blueshifted He~II~$\lambda$4686 line, formed from the SN ejecta beneath a CSM shell \citep{bullivant18,andrews19}, although when this solution was proposed, there was a narrow component of this line present. This interpretation, however, is hard to reconcile with the irregular profile of the feature. Alternatively, the ledge might result from a blend of several high-ionization lines of C, N, and O originating in the CSM \citep{soumagnac20,bruch21}. However, this hypothesis is complicated by the feature's width, as these high-ionization lines would need to be significantly broader or more numerous to explain the observed profile. \citet{hosseinzadeh22} offered another possibility, suggesting that the explosion of an RSG with an extended envelope into a low-density CSM could produce the ledge-shaped feature without generating the high-ionization lines typically associated with flash ionization. The closer association of ledger SNe with the non-flash group would suggest that the origin of the ledge feature may be different from that of the flash features.

\section{Conclusions}
In this work we have presented a sample of 68 SNe~II with early-time photometric observations from the ATLAS survey and complementary spectroscopy from the ePESSTO+ collaboration. By combining well-constrained light curves with early and photospheric spectra, we investigated the impact of CSM interaction on the photometric and spectroscopic evolution of SNe II.

We identified early-time high-ionization, electron-scattered broadened spectral features ---a sign of CSM interaction--- in 18 SNe (flash SNe), and contrasted their properties with those of 25 non-flash SNe. An additional subset of 8 SNe displayed the ambiguous ``ledge'' feature, and we evaluated their properties under both the flash and non-flash assumptions to test their association.

Our main findings are as follows:
\begin{itemize}
    \item Flash SNe exhibit significantly faster $s_{1}$ decline rates and higher peak luminosities compared to non-flash SNe. This suggests that early interaction with CSM not only affects spectral features but also alters the energy release observed in the light curves. However, no significant difference in rise time was found between flash and non-flash SNe, maybe suggesting that rise times alone are not a reliable diagnostic for early interaction.
    \item A clear distinction in the $\mathrm{H\alpha}$ a/e ratio was observed between the two groups. Flash SNe consistently show lower a/e values at all epochs, which may reflect suppression of absorption due to enhanced CSM interaction or altered density structures in the outer ejecta.
    \item SNe displaying the ledge feature tend to align more closely with the non-flash population in terms of decline rates, peak brightness, and a/e ratios. While the origin of this feature remains uncertain, our results suggest that these events may represent a lower level or different manifestation of interaction, distinct from that seen in flash SNe.
    \item A subset of SNe II in our sample, without considering the 87A-like events, exhibits long rise times (>15 days). These include both flash and non-flash SNe, suggesting multiple physical origins. Some may arise from extended progenitor envelopes or dense CSM capable of delaying shock breakout, while others may reflect standard RSG explosions with minimal surrounding material.
\end{itemize}

This work shows the value of combining early-time photometric and spectroscopic observations to characterise the diversity of SNe II and the frequency and impact of early CSM interaction. Future high-cadence surveys with rapid spectroscopic follow-up will be key to improving our understanding of the final stages of massive star evolution and the role of CSM in shaping SN observables.

\begin{acknowledgements}
This work has made use of data from the Asteroid Terrestrial-impact Last Alert System (ATLAS) project. The Asteroid Terrestrial-impact Last Alert System (ATLAS) project is primarily funded to search for near earth asteroids through NASA grants NN12AR55G, 80NSSC18K0284, and 80NSSC18K1575; byproducts of the NEO search include images and catalogs from the survey area. This work was partially funded by Kepler/K2 grant J1944/80NSSC19K0112 and HST GO-15889, and STFC grants ST/T000198/1 and ST/S006109/1. The ATLAS science products have been made possible through the contributions of the University of Hawaii Institute for Astronomy, the Queen’s University Belfast, the Space Telescope Science Institute, the South African Astronomical Observatory, and The Millennium Institute of Astrophysics (MAS), Chile. 
Based on observations collected at the European Southern Observatory under the ePESSTO+ programmes with ID: 106.216C, 108.220C, and 1103.D-0328 (PI: Inserra). CPG acknowledges financial support from the Secretary of Universities and Research (Government of Catalonia) and by the Horizon 2020 Research and Innovation Programme of the European Union under the Marie Sk\l{}odowska-Curie and the Beatriu de Pin\'os 2021 BP 00168 programme, from the Spanish Ministerio de Ciencia e Innovaci\'on (MCIN) and the Agencia Estatal de Investigaci\'on (AEI) 10.13039/501100011033 under the PID2023-151307NB-I00 SNNEXT project, from Centro Superior de
Investigaciones Científicas (CSIC) under the PIE project 20215AT016 and the program Unidad de Excelencia María de Maeztu CEX2020-001058-M, and from the Departament de Recerca i Universitats de la Generalitat de Catalunya through the 2021-SGR-01270 grant.

AA acknowledges the Yushan Young Fellow Program by the Ministry of Education, Taiwan, for the financial support (MOE-111-YSFMS-0008-001-P1). T.-W.C. acknowledges the financial support from the Yushan Fellow Program by the Ministry of Education, Taiwan (MOE-111-YSFMS-0008-001-P1) and the National Science and Technology Council, Taiwan (NSTC grant 114-2112-M-008-021-MY3). TP and EC acknowledges the financial support from the Slovenian Research Agency (grants I0-0033, P1-0031, J1-8136, J1-2460 and the Young researchers program). L.G. acknowledges financial support from AGAUR, CSIC, MCIN and AEI 10.13039/501100011033 under projects PID2023-151307NB-I00, PIE 20215AT016, CEX2020-001058-M, ILINK23001, COOPB2304, and 2021-SGR-01270. T.E.M.B. is funded by Horizon Europe ERC grant no. 101125877.

\end{acknowledgements}
\bibliographystyle{aa} 
\bibliography{typeIIs} 

\onecolumn
\begin{appendix}
\section{Properties of the sample}
\label{ap:1}
\begin{longtable}{lcccccc}
  \caption{SN II Sample}
  \label{tab:sample}\\
  \hline \multicolumn{1}{c}{SN} & \multicolumn{1}{c}{Host Galaxy} & \multicolumn{1}{c}{z} & \multicolumn{1}{c}{$E(B-V)_{MW}$} & \multicolumn{1}{c}{\CellWithForecedBreak{Last \\ non-det}}&\multicolumn{1}{c}{\CellWithForecedBreak{Discovery \\ Date}} & \multicolumn{1}{c}{\CellWithForecedBreak{Exp \\ epoch}}   \\ \hline 
  \endfirsthead
  \multicolumn{3}{c}%
{{\tablename\ \thetable{} - (continued)}} \\
\hline \multicolumn{1}{c}{SN} & \multicolumn{1}{c}{Host Galaxy} & \multicolumn{1}{c}{z} & \multicolumn{1}{c}{$E(B-V)_{MW}$} & \multicolumn{1}{c}{\CellWithForecedBreak{Last \\ non-det}}& \multicolumn{1}{c}{\CellWithForecedBreak{Discovery \\ Date}} & \multicolumn{1}{c}{\CellWithForecedBreak{Exp \\ epoch}}  \\ \hline
\endhead
  SN~2019ofc  & LCRS B013124.4-385126        & 0.020 & 0.01  & 58715.63	&58717.51	&58716.57 \\
  SN~2019pnl  & LEDA 997304                  & 0.040 & 0.07  & 58726.56	&58730.55	&58728.55 \\
  SN~2019ssi  & UGC 12640                    & 0.013 & 0.06  & 58773.38	&58775.37   &58774.37  \\
  SN~2019uhm  & WISEA J102849.18-312953.3    & 0.010 & 0.05  & 58789.64	&58791.64	&58790.64 \\
  SN~2019vjl  & NGC 3015                     & 0.025 & 0.07  & 58802.64	&58808.62	&58805.63 \\
  SN~2019xtw  & ESO299-G013                  & 0.016 & 0.01  & 58837.34	&58839.27	&58838.30 \\
  SN~2020aze  & NGC 3318                     & 0.009 & 0.06  & 58871.50	&58875.47	&58873.48 \\
  SN~2020bij  & NGC 3463                     & 0.013 & 0.06  & 58875.56	&58877.51	&58876.53 \\
  SN~2020szs  & CGCG 402-005                 & 0.022 & 0.03  & 59102.39	&59103.40	&59102.90 \\
  SN~2020vef  & SDSS J034211.73-002113.0     & 0.026 & 0.06  & 59130.52	&59132.49	&59131.51 \\
  SN~2020voh  & 6dFJ230350.8-360124 (LARS)   & 0.016 & 0.001 & 59129.38	&59131.37	&59130.37 \\
  SN~2020xoq  & 2MASX J20030114-4133024      & 0.031 & 0.08  & 59137.24	&59139.22	&59138.23 \\
  SN~2020znl  & PS1 88920933245169840        & 0.037 & 0.23  & 59160.54	&59164.51	&59162.52 \\
  SN~2020abyj & ESO476-G023                  & 0.032 & 0.01  & 59189.32	&59191.29	&59190.31 \\
  SN~2020acbm & LSBC F831-08                 & 0.022 & 0.02  & 59192.40	&59194.35	&59193.38 \\
  SN~2021aek  & I SZ 091                     & 0.022 & 0.04  & 59223.59	&59227.57	&59225.58 \\
  SN~2021cgu  & CGCG 038-099                 & 0.025 & 0.03  & 59248.57	&59252.52	&59250.54  \\
  SN~2021dbg  &S4YI000212 (GSC cat)          & 0.020 & 0.02  & 59256.44	&59260.42	&59258.43 \\
  SN~2021gvm  & SDSS J132959.99+132446.0     & 0.023 & 0.02  & 59293.50	&59297.42	&59295.46 \\
  SN~2021mqh  & MCG -01-30-021               & 0.021 & 0.03  & 59350.39	&59352.31	&59351.35 \\
  SN~2021qvr  & NGC 7678                     & 0.011 & 0.04  & 59383.57	&59387.60   &59385.58 \\
  SN~2021rem  & 2MASX J15194792+1341104      & 0.029 & 0.03  & 59391.36	&59392.38	&59391.87 \\
  SN~2021tiq  & MCG -02-57-024               & 0.023 & 0.05  & 59408.51	&59412.51	&59410.51 \\
  SN~2021tyw  & UGC 12354                    & 0.013 & 0.20  & 59417.62	&59419.50	&59418.56  \\
  SN~2021tsz  & SDSS J233758.38-002629.6     & 0.040 & 0.03  & 59412.57	&59414.51	&59413.54 \\
  SN~2021sqg  & ESO287-G018                  & 0.015 & 0.02  & 59399.62	&59403.46	&59401.54 \\
  SN~2021wvd  & CGCG077-028                  & 0.045 & 0.03  & 59449.31	&59451.27	&59450.29 \\
  SN~2021yyg  & 2MASX J05162092-1328164      & 0.011 & 0.13  & 59470.62	&59471.59	&59471.11 \\
  SN~2021wyn  & 2MASS J00411206-1442024      & 0.053 & 0.02  & 59445.46	&59447.45	&59446.46 \\
  SN~2021aatd & GALEXASC J005904.40-001210.0 & 0.015 & 0.02  & 59492.51	&59494.37	&59493.44 \\
  SN~2021adcw & 2MASS J02043283-2107094      & 0.018 & 0.01  & 59519.49	&59523.44	&59521.46 \\
  SN~2021afkk & UGC 01971                    & 0.015 & 0.09  & 59541.42	&59543.37   &59542.39 \\
  SN~2021agbn & 2MASX 03571315-2523551       & 0.014 & 0.03  & 59547.43	&59551.36   &59549.40 \\
  SN~2021adyl & IC2479                       & 0.028 & 0.02  & 59523.61	&59525.58	&59524.59 \\
  SN~2021yja  & NGC1325                      & 0.005 & 0.01  & 59463.48	&59465.55	&59464.51 \\
  SN~2022bcf  & MCG -01-12-043               & 0.016 & 0.10  & 59608.84	&59610.24	&59609.54 \\
  SN~2022jo   & IC 4040                      & 0.025 & 0.009 & 59587.62	&59589.46	&59588.54 \\
  SN~2022mm   & UGCA258                      & 0.013 & 0.04  & 59588.57	&59590.64	&59589.60 \\
  SN~2022abq  & NGC 5117                     & 0.008 & 0.01  & 59597.55	&59601.59	&59599.57 \\
  SN~2022big  & ORPHAN                       & 0.018 & 0.017 & 59604.88	&59606.87	&59605.88 \\
  SN~2022fqe  & NGC 5995                     & 0.025 & 0.14  & 59669.06	&59670.53	&59669.79 \\
  SN~2022etp  & 2MASX J14221000+2505490      & 0.038 & 0.01  & 59649.49	&59651.48	&59650.49 \\
  SN~2022kad  & CGCG076-118                  & 0.019 & 0.03  & 59712.51	&59716.49	&59714.50  \\
  SN~2022jdf  & SDSS J151828.21+051422.0     & 0.035 & 0.04  & 59699.22	&59700.40	&59699.81 \\
  SN~2022iek  & UGC05818                     & 0.021 & 0.02  & 59691.12	&59692.33	&59691.73 \\
  SN~2022ibv  & ESO437-G064                  & 0.014 & 0.05  & 59690.15	&59690.87	&59690.51 \\
  SN~2022frq  & MCG -02-34-054               & 0.022 & 0.05  & 59670.43	&59672.92	&59671.68 \\
  SN~2022fli  & HIDEEP J1342-3405            & 0.025 & 0.04  & 59661.33	&59665.90	&59664.12 \\
  SN~2022jox  & ESO 435- G 014               & 0.008 & 0.09  & 59706.04	&59709.02	&59707.53 \\
  SN~2022ffg  & CGCG 093-074                 & 0.012 & 0.035 & 59663.35	&59664.30	&59663.83 \\
  SN~2022rje  & GALEXASCJ045753.84-161502.7  & 0.021 & 0.076 & 59806.62	&59807.39	&59807.01 \\
  SN~2022yjl  & SDSS J221758.36+105943.4     & 0.080 & 0.059 & 59872.82	&59874.34	&59873.58 \\
  SN~2022xkp  & LCRS B025030.8-412847        & 0.012 & 0.016 & 59860.18	&59862.28	&59861.73 \\
  SN~2022xko  & ESO 014- G 001               & 0.015 & 0.076 & 59861.98	&59864.27	&59863.12 \\
  SN~2022wsp  & NGC7448                      & 0.007 & 0.04  & 59854.40	&59855.13	&59854.77 \\
  SN~2022zkc  & GIN227                       & 0.032 & 0.04  & 59884.03	&59885.20   &59884.62 \\
  SN~2022aagp & NGC 2777                     & 0.005 & 0.038 & 59894.57	&59897.55	&59896.06 \\
  SN~2022aacn & 	LEDA 5807554               & 0.018 & 0.031 & 59895.43	&59896.39	&59895.91  \\
  SN~2022acmt & ESO547-G027                  & 0.033 & 0.03  & 59921.18	&59922.14	&59921.66  \\
  SN~2022acko & 2MASXiJ0319359-192344        & 0.006 & 0.02  & 59918.17	&59919.92	&59919.04  \\
  SN~2022acug & S5XX016714 (GSC cat)         & 0.010 & 0.078 & 59921.91	&59924.15	&59923.03  \\
  SN~2022acyd & SDSS J040111.13+053305.7     & 0.035 & 0.237 & 59927.16	&59928.69	&59927.93  \\
  SN~2022acrv & PS1 83820866869460987        & 0.034 & 0.045 & 59921.27	&59922.25	&59921.76    \\
  SN~2023cr   & ESO419-G003                  & 0.014 & 0.011 & 59950.90	&59951.83	&59951.36  \\
  SN~2023dr   & VIII Zw 154                  & 0.035 & 0.042 & 59951.29	&59952.97	&59952.13 \\
  SN~2023ng   & 2MASX gJ091522.0-185217      & 0.017 & 0.058 & 59957.25	&59958.29	&59957.77 \\
  SN~2023aub  & IC 4219                      & 0.012 & 0.057 & 59969.37	&59970.31   &59969.84 \\
  SN~2023azx  & PS1 88791808263516143        & 0.011 & 0.045 & 59971.32	&59972.05	&59971.69 \\
  \hline
\end{longtable}

\begin{longtable}{lcccc}
  \caption{SN II Sample spectral information}
  \label{tab:sample2}\\
  \hline \multicolumn{1}{c}{SN} & \multicolumn{1}{c}{\CellWithForecedBreak{N of \\ spectra}} & \multicolumn{1}{c}{\CellWithForecedBreak{Earliest and \\ latest epoch}}  & \multicolumn{1}{c}{Classification} & \multicolumn{1}{c}{\CellWithForecedBreak{Reference for \\ classification}} \\ \hline 
  \endfirsthead
  \multicolumn{3}{c}%
{{\tablename\ \thetable{} - (continued)}} \\
\hline \multicolumn{1}{c}{SN} & \multicolumn{1}{c}{\CellWithForecedBreak{N of \\ spectra}} & \multicolumn{1}{c}{\CellWithForecedBreak{Earliest and \\ latest epoch}}  & \multicolumn{1}{c}{Classification} & \multicolumn{1}{c}{\CellWithForecedBreak{Reference for \\ classification}} \\ \hline 
\endhead
  SN~2019ofc  & 5  & 1.7 - 107    &  F   &  This work  \\
  SN~2019pnl  & 4  & 49.7 - 94.5  &  F   &  \citet{fremling19}  \\
  SN~2019ssi  & 3  & 2.7 - 51.7   &  NF  &  This work \\
  SN~2019uhm  & 5  & 16.7 - 113.5 &  N/A &  -  \\
  SN~2019vjl  & 5  & 10.7 - 98.5  &  N/A &  -  \\
  SN~2019xtw  & 1  & 42.8         &  NF  &  \citet{do20}  \\
  SN~2020aze  & 4  & 7.7 - 31.7   &  NF  &  This work \\
  SN~2020bij  & 2  & 4.7 - 46.7   &  NF  &  This work   \\
  SN~2020szs  & 1  & 60.2         &  NF  &  \citet{srivastav20}  \\
  SN~2020vef  & 6  & 56.7 - 116.6 &  NF  &  \citet{dahiwale20} \\
  SN~2020voh  & 4  & 15.8 - 87.7  &  N/A &  -  \\
  SN~2020xoq  & 2  & 9.8 - 23.9   &  NF  &  This work  \\
  SN~2020znl  & 9  & 11.6 - 85.6  &  N/A &  -  \\
  SN~2020abyj & 5  & 3.8 - 59.7   &  F   &  This work  \\
  SN~2020acbm & 4  & 4.8 - 56.7   &  NF  &  This work   \\
  SN~2021aek  & 9  & 3.8 - 61.8   &  F   &  This work  \\
  SN~2021cgu  & 3  & 5.7 - 106.5  &  F   &  This work  \\
  SN~2021dbg  & 3  & 1.7 - 54.6   &  F   &  This work   \\
  SN~2021gvm  & 3  & 0.8 - 24.7   &  NF  &  This work   \\
  SN~2021mqh  & 0  & -            &  F   &  \citet{jacobsongalan24}  \\
  SN~2021qvr  & 4  & 42.8 - 106.6 &  F   &  \citet{jacobsongalan24}  \\
  SN~2021rem  & 0  & -            &  NF  &  \citet{dimitriadis21}  \\
  SN~2021tiq  & 8  & 17.7 - 114.6 &  NF  &  \citet{srivastav21}  \\
  SN~2021tyw  & 18 & 6.7 - 110.5  &  F   &  This work  \\
  SN~2021tsz  & 1  & 14.8         &  N/A &  -  \\
  SN~2021sqg  & 4  & 30.6 - 90.5  &  N/A &  -  \\
  SN~2021wvd  & 4  & 2.7 - 9.7    &  F   &  This work   \\
  SN~2021yyg  & 7  & 12.3 - 89.2  &  NF  &  \citet{burke21}  \\
  SN~2021wyn  & 2  & 54.7         &  N/A &  -  \\
  SN~2021aatd & 3  & 8.8 - 30.8   &  NF  &  This work  \\
  SN~2021adcw & 3  & 21.8 - 59.7  &  N/A &  -  \\
  SN~2021afkk & 0  & -            &  F   &  \citet{jacobsongalan24}   \\
  SN~2021agbn & 3  & 2.7 - 110.6  &  L   &  This work   \\
  SN~2021adyl & 1  & 49.7         &  N/A &  - \\
  SN~2021yja  & 5  & 6.8 - 16.8   &  L   &  This work  \\
  SN~2022bcf  & 1  & 3.6          &  NF  &  This work   \\
  SN~2022jo   & 1  & 62.8         &  NF  &  \citet{li22c}  \\
  SN~2022mm   & 4  & 2.7 - 69.7   &  NF  &  This work    \\
  SN~2022abq  & 1  & 50.7         &  NF  &  \citet{ochner22}  \\
  SN~2022big  & 0  & -            &  N/A &   - \\
  SN~2022fqe  & 0  & -            &  NF  &  \citet{fulton22} \\
  SN~2022etp  & 0  & -            &  NF  &  \citet{payne22} \\
  SN~2022kad  & 0  & -            &  NF  &  \citet{hinkle22b}  \\
  SN~2022jdf  & 1  & 22.5         &  N/A &  -  \\
  SN~2022iek  & 1  & 11.4         &  F   &  This work  \\
  SN~2022ibv  & 0  & -            &  F   &  \citet{jacobsongalan24}  \\
  SN~2022frq  & 0  & -            &  NF  &  \citet{jacobsongalan24}   \\
  SN~2022fli  & 1  & 39.1         &  N/A &  -  \\
  SN~2022jox  & 0  & -            &  F   &  \citet{jacobsongalan24}   \\
  SN~2022ffg  & 3  & 4.3 - 7.3    &  F   &  This work   \\
  SN~2022rje  & 1  & 36.3         &  N/A &  -  \\
  SN~2022yjl  & 0  & -            &  N/A &  -  \\
  SN~2022xkp  & 3  & 5.5 - 81.4   &  NF  &  This work  \\
  SN~2022xko  & 2  & 3.0 - 37.2   &  NF  &  This work   \\
  SN~2022wsp  & 1  & 61.3         &  NF  &  \citet{nagao22}  \\
  SN~2022zkc  & 1  & 71.5         &  N/A &  -  \\
  SN~2022aagp & 4  & 48.3 - 88.2  &  L   &  \citet{taguchi22}  \\
  SN~2022aacn & 0  & -            &  NF  &  \citet{hinkle22}  \\
  SN~2022acmt & 1  & 52.4         &  L   &  \citet{terreran22}  \\
  SN~2022acko & 2  & 18.1 - 39.0  &  L   &  \citet{lin25}  \\
  SN~2022acug & 3  & 12.3 - 61.3  &  N/A &  - \\
  SN~2022acyd & 1  & 56.1         &  F   &  \citet{li22}  \\
  SN~2022acrv & 2  & 22.4 - 36.3  &  N/A &  -  \\
  SN~2023cr   & 2  & 6.7 - 32.7   &  F   &  This work  \\
  SN~2023dr   & 0  & -            &  F   &  \citet{hinkle23}  \\
  SN~2023ng   & 2  & 1.5 - 33.3   &  L   &  This work   \\
  SN~2023aub  & 0  & -            &  L   &  \citet{pellegrino23}  \\
  SN~2023azx  & 2  & 2.7 - 24.6   &  L   &  This work   \\
  \hline
\end{longtable}

\begin{table*}[ht!]
\centering
	\caption{LC properties compared with previous works in the literature. When available, the standard error of the mean is listed.}
	\label{tab:LCdist}
	\begin{tabular}{lcccc} 
		\hline
		Reference & N of SNe & Wavelength [{\AA}] / band & $\mathrm{t_{rise}}$~[days] & $\mathrm{M_{max}}$~[mag]  \\
		\hline
            \multirow{2}{*}{This work} & 64 & $o$ & $11.8 \pm 0.8$ & $-17.7 \pm 0.1$  \\
            & 21 & $c$ &$10.7 \pm 1.7$&$-18.0 \pm 0.2$\\ 
            \citet{tammann90} & 23 & $B$ & - & $-17.2 \pm 0.3$ \\ 
            \citet{anderson14} & 116 & $V$ & - & $-16.7 \pm 0.1$ \\ 
            \citet{gutierrez14} & 52 & $V$ & & $-16.7$ \\ 
            \multirow{2}{*}{\citet{gonzalezgaitan15}} & 223 & <4000 & <10 & - \\ 
            & 223 & >7000 & 12-15 & - \\ 
		\citet{gall15} & 20 & $r'$/$R$ & $10.2 \pm 0.2$ & - \\ 
            \citet{rubin16} & 57 & $R$ & $7.7 \pm 0.4$ \\ %
            \citet{gutierrez17b} & 99 & $V$ & - & $-16.7 \pm 0.1$\\ 
            \multirow{3}{*}{\citet{pessi19}} & 73 & $B$ & $8.3 \pm 0.2$ & - \\ 
		& 73 & $V$ & $12.8 \pm 0.3$ & - \\ 
		& 73 & $r$ & $16 \pm 0.4$ & - \\ 
            \multirow{2}{*}{\citet{bruch21}} & 17 &$g$ & $8.9$ & $-17.7$ \\ 
            & 16 &$r$ & 12.7 & -17.7\\ 
            \citet{rodriguez21} & 60 & $R$ & - & $-16.7 \pm 0.1$ \\ 
            \citet{das25} & 129 & $r$ & $13^{+0.4}_{-0.2}$ & - \\ 
    	\hline
	\end{tabular}
\end{table*}

\section{SNe light-curves with their fits}
\label{ap:2}
\begin{figure*}[!ht]
\centering
\includegraphics[width=1.0\textwidth]{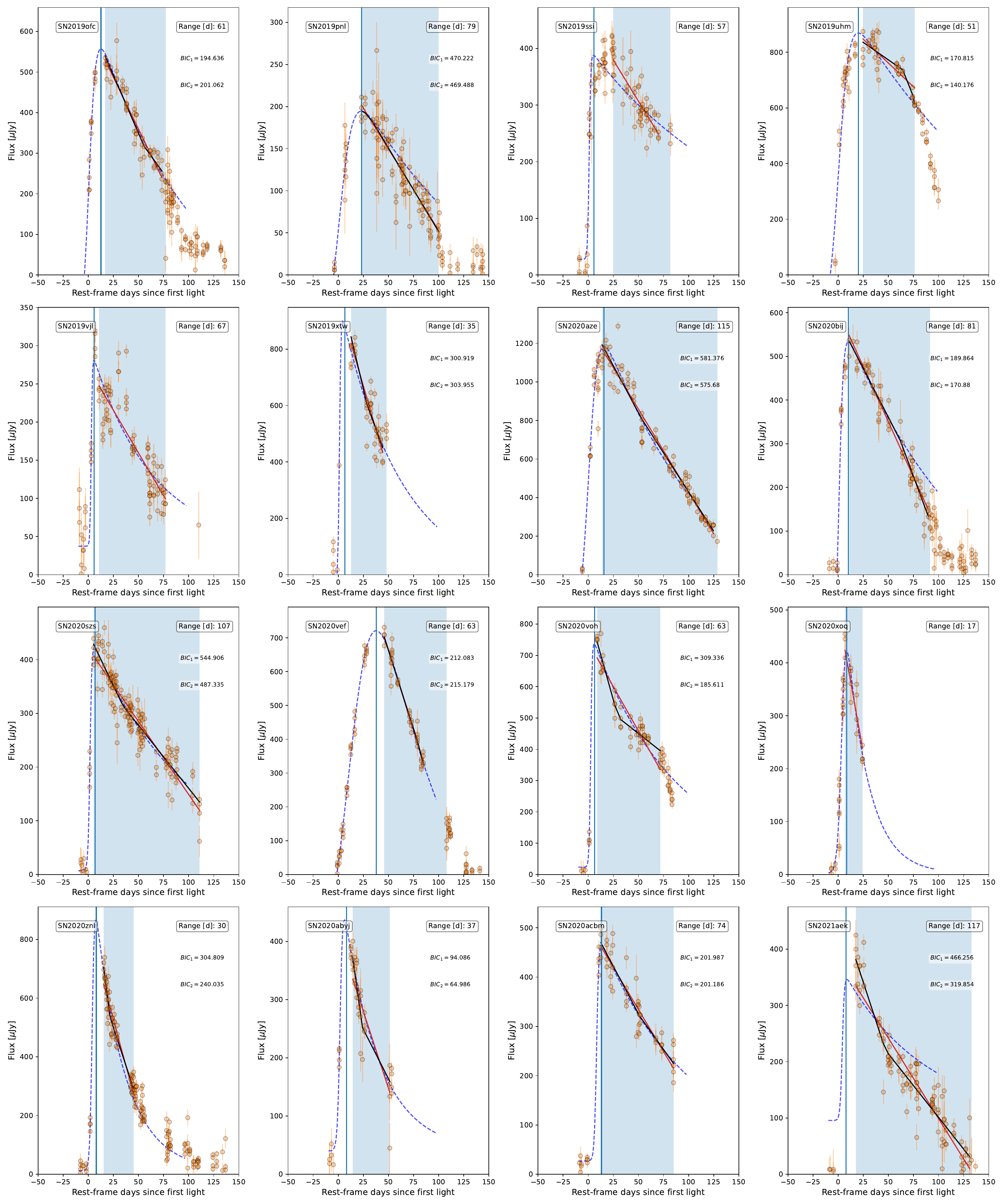}
\caption{LCs of the SNe~II from the sample in the $o$ band shown in orange. Marked with dashed blue lines are the fits to the LCs. The time of maximum light is shown with vertical lines. The light blue bands mark the range where the fits to calculate the decline rates were made. Marked with red lines are the fits to the LCs with one slope and with black are the fits with two slopes. In those cases where the two slopes regime provided bad fits, only the fits with one slope are plotted.}
\label{lcs1}
\end{figure*}

\begin{figure*}[!ht]
\ContinuedFloat
\includegraphics[width=1.0\textwidth]{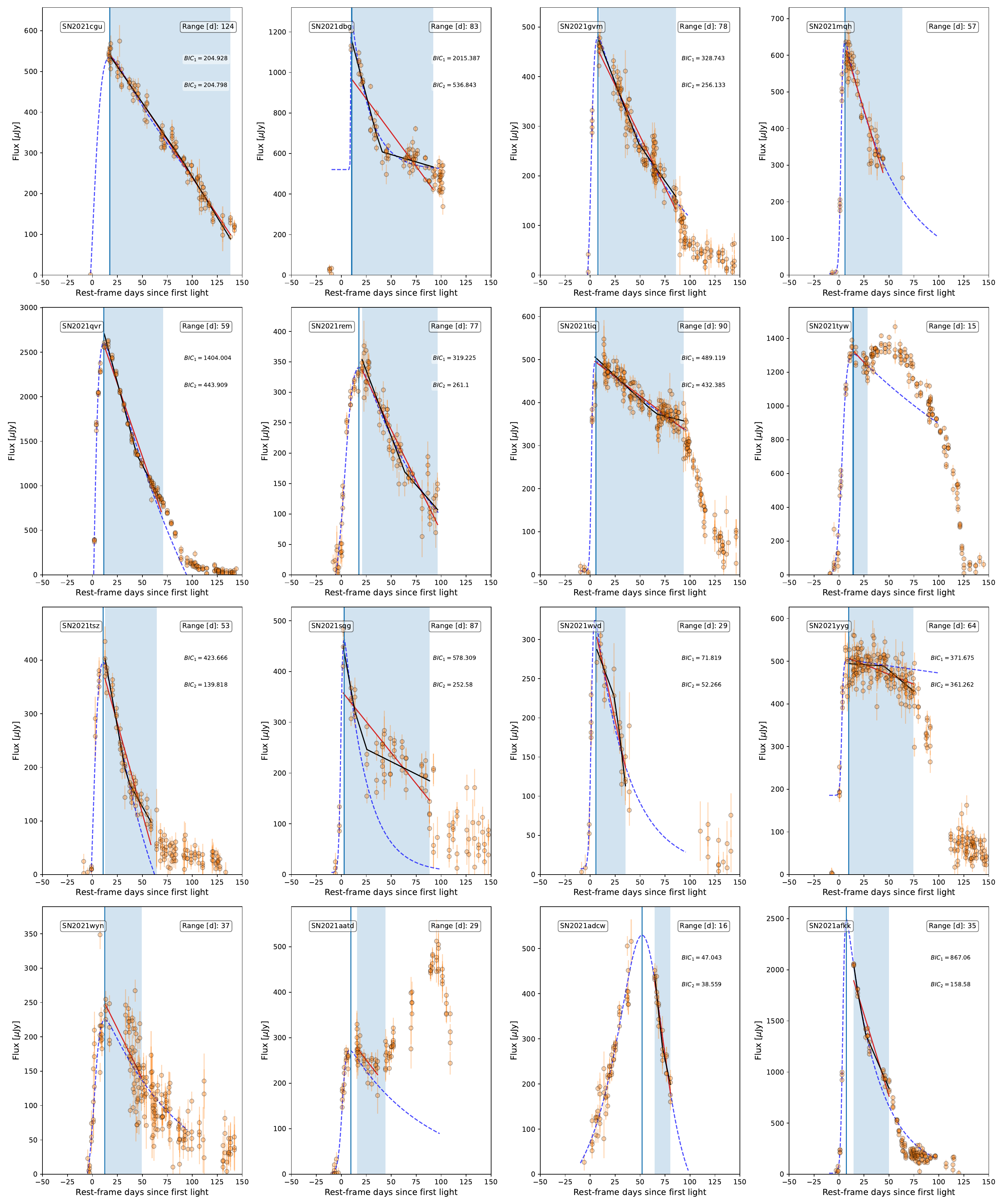}
\caption{Continued.}
\label{lcs2}
\end{figure*}

\textbf{\begin{figure*}[!ht]
\ContinuedFloat
\includegraphics[width=1.0\textwidth]{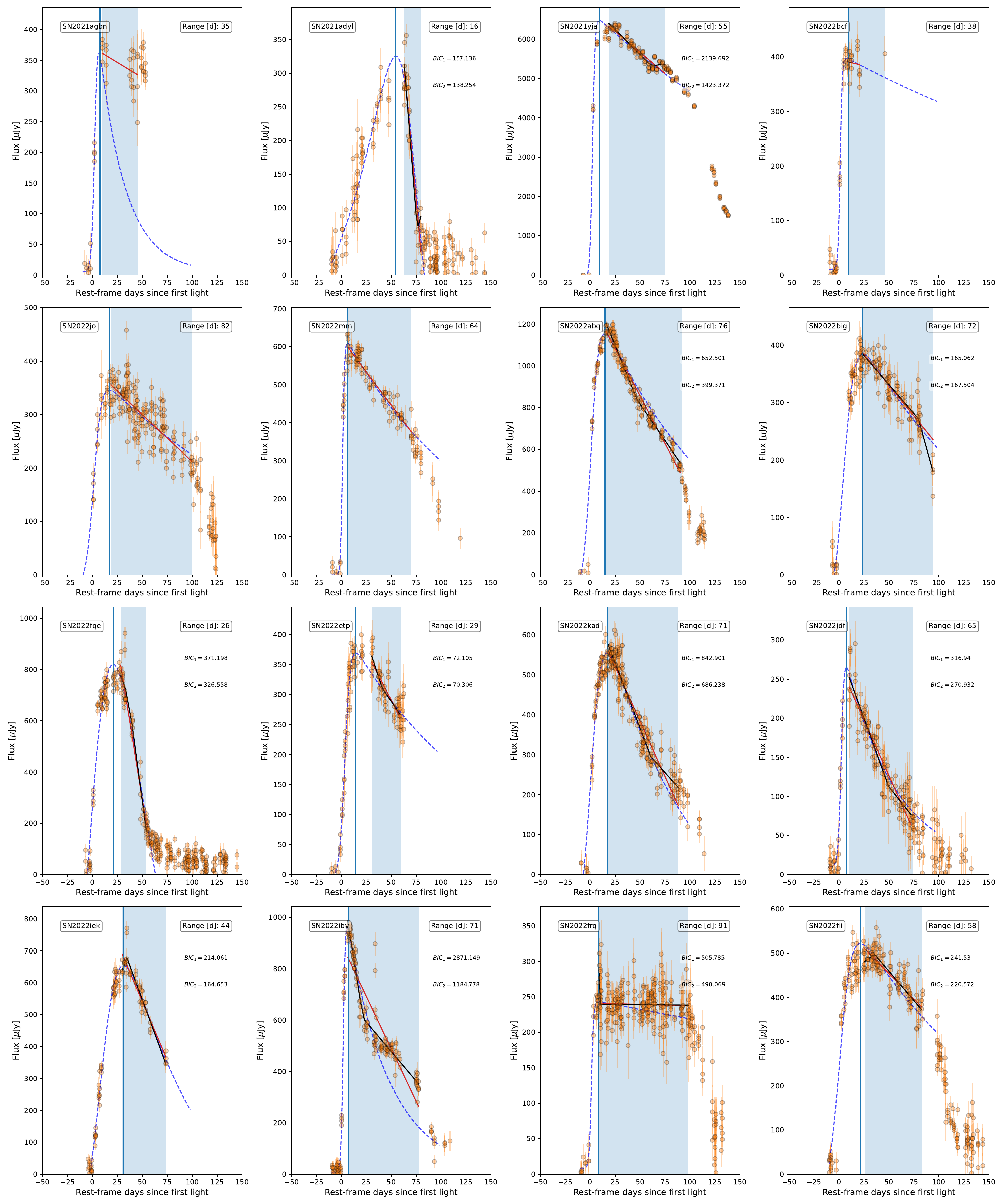}
\caption{Continued.}
\label{lcs3}
\end{figure*}}

\begin{figure*}[!ht]
\ContinuedFloat
\centering
\includegraphics[width=0.8\textwidth]{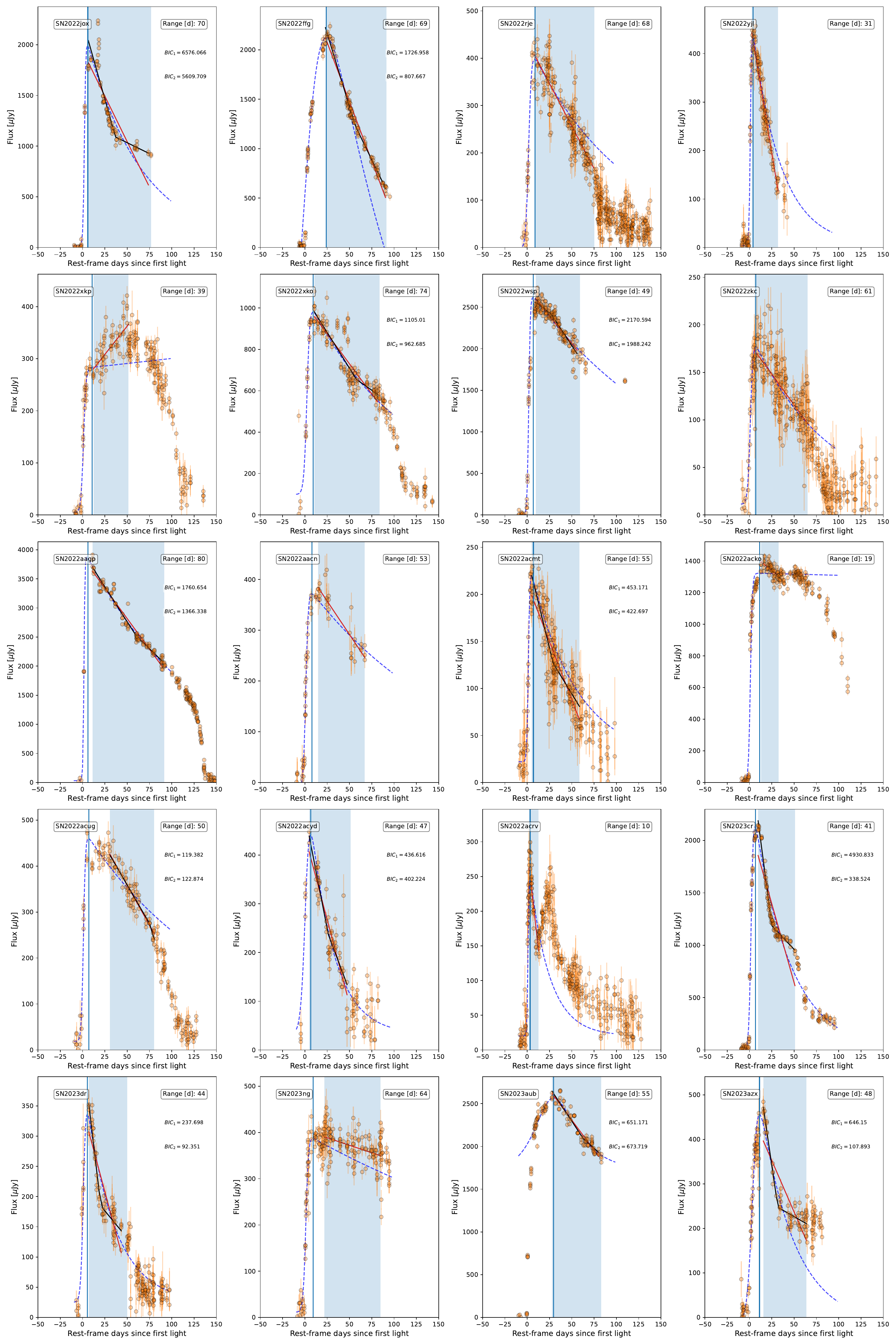}
\caption{Continued.}
\label{lcs4}
\end{figure*}

\begin{figure*}[!ht]
\centering
\includegraphics[width=0.78\textwidth]{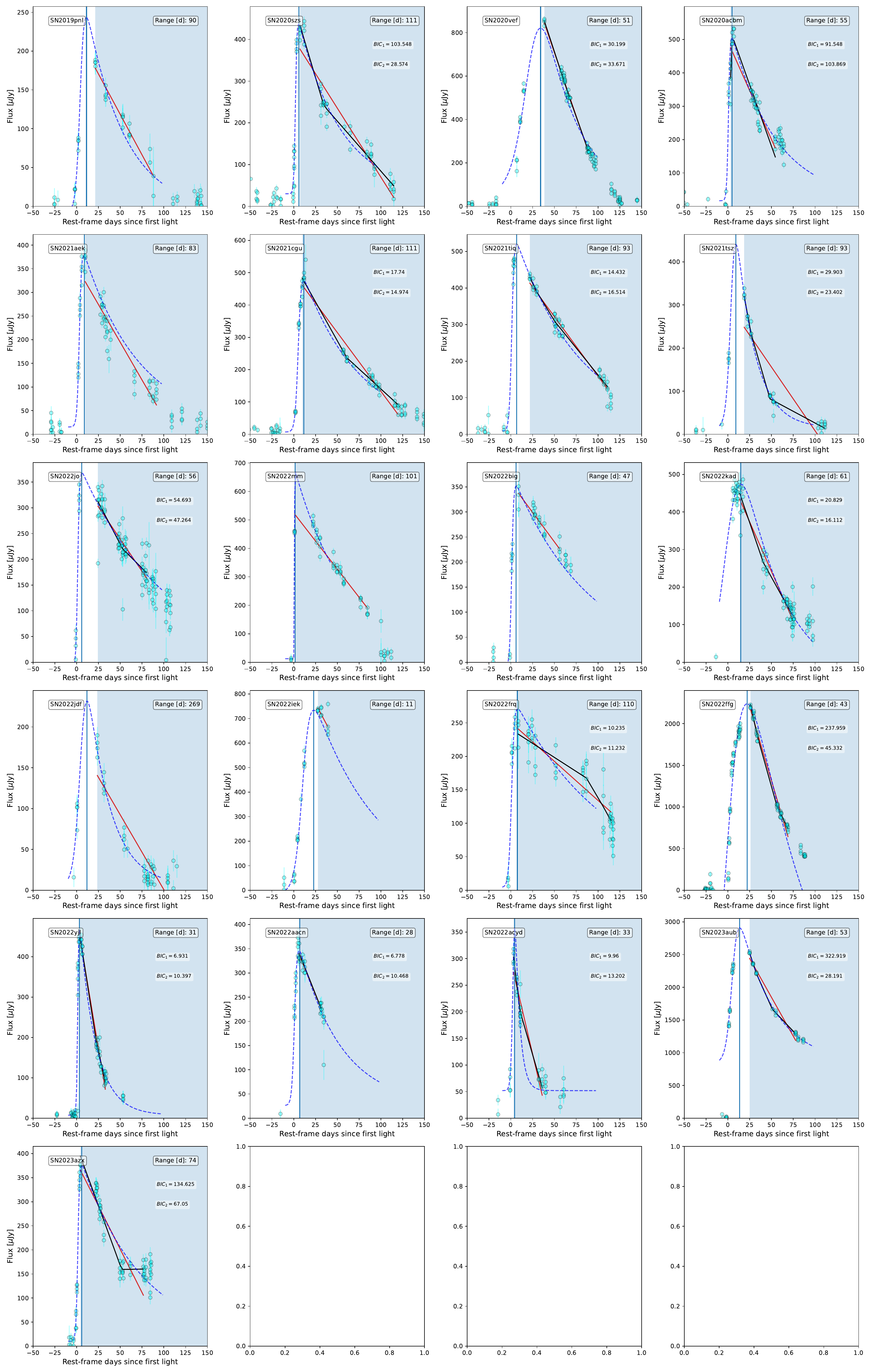}
\caption{LCs of the SNe~II from the sample in the $c$ band shown in cyan. Marked with dashed bluelines are the fits to the LCs. The light blue bands mark the range where the fits to calculate the decline rates were made. Marked with red lines are the fits to the LCs with one slope and with black are the fits with two slopes. The time of maximum light is shown with vertical lines. In those cases where the two slopes regime provided bad fits, only the fits with one slope are plotted. }
\label{dec_cband}
\end{figure*}

\end{appendix}
\label{LastPage}
\end{document}